\documentclass[usenatbib]{mn2e}

\newcommand\dosingle[1]{#1}  \newcommand\dodouble[1]{ } 

\newcommand\nice[1]{#1}    \newcommand\subm[1]{}   
\subm{ \renewcommand\dosingle[1]{ }   \renewcommand\dodouble[1]{#1} }

\dodouble{\documentclass[usenatbib,referee]{mn2e}}

\usepackage{graphicx}

\usepackage{amsfonts}
\usepackage{amsmath}

\usepackage{mn_bmath} 

\usepackage[varg]{txfonts}

\newcommand\mystamp[1]{{ }}

\newcommand\mystamppreamble{
  \usepackage{eso-pic}
  \usepackage{color} 
  \definecolor{redstamp}{rgb}{0.99,0.80,0.90} 
  \usepackage{datetime}
  \usepackage[normalem]{ulem}
}

\newcommand\mystampdothestamp{
  \AddToShipoutPicture{
    \AtTextLowerLeft{
      \makebox(550,300)[c]{\resizebox{\textwidth}{!}{
          \rotatebox{36}{\textsf{\textbf{\color{redstamp}brouillon~\frdutoday~\currenttime}}}}} 
    }
  }
}

\mystamp{\mystamppreamble}

\usepackage{natbib}  
\usepackage{url} 

\newcommand\postrefereechanges[1]{#1}     \newcommand\postrefereetitle[1]{#1}

\newcommand\postrefereechangesbis[1]{#1}

 \usepackage{hyperref}

\nice{ 
\providecommand{\eprint}[1]{\href{http://arxiv.org/abs/#1}{{\tt [arXiv:#1]}}}

\providecommand{\url}[1]{\href{#1}{#1}}

}

\providecommand{\adsurl}[1]{} 
\newcommand\SSS{Sect.~}

\providecommand\apj{ApJ}                 
\providecommand\aj{AJ}                 
\providecommand\apjs{ApJSupp}                 
\providecommand\apjl{ApJL}                 
\providecommand\aap{A\&A}            
\providecommand\mnras{MNRAS}

\providecommand\PRL{Physical Review Letters}

\providecommand\PRL{PRL}
\providecommand\prd{PRD}

\providecommand\jcap{JCAP}

\providecommand\pasj{PASJ}

\providecommand\grg{Gen. Rel. Grav.}
\providecommand\ijmpd{Int. J. Mod. Phys. D}
\providecommand\jsiam{J.~Soc.~Indust.~App.~Math.}

\nice{ \newcommand\mycaptionfont{\protect\footnotesize} }

\newcommand\gtapprox{\,\lower.6ex\hbox{$\buildrel >\over \sim$} \, }
\newcommand\ltapprox{\,\lower.6ex\hbox{$\buildrel <\over \sim$} \, }
\newcommand\propapprox{\,\lower.6ex\hbox{$\buildrel \propto\over \sim$} \, }

\newcommand\arcs{\ifmmode {'' }\else $'' $\fi}     
\newcommand\arcm{\ifmmode {' }\else $' $\fi}       
\newcommand\ddeg{\ifmmode^\circ\else$^\circ$\fi}    

\newcommand\frtoday{Le\space\number\day\space\ifcase\month\or
  janvier\or f\'evrier\or mars\or avril\or mai\or juin\or
  juillet\or ao\^ut\or septembre\or octobre\or novembre\or 
d\'ecembre\fi\space \number\year}
\def\frdutoday{du\space\number\day\space\ifcase\month\or
  janvier\or f\'evrier\or mars\or avril\or mai\or juin\or
  juillet\or ao\^ut\or septembre\or octobre\or novembre\or 
d\'ecembre\fi\space \number\year}

\newcommand\todayISO{\number\year-\ifnum\month<10 0\fi\number\month-\ifnum\day<10 0\fi\number\day}

\newcommand\cqg{ClassQuantGra}   %
\newcommand\hGpc{\mbox{$h^{-1}$ Gpc}}
\newcommand\hMpc{\mbox{$h^{-1}$ Mpc}}
\newcommand\hMpcmath{h^{-1}\mathrm{Mpc}}

\newcommand\Ommzero{\Omega_{\mathrm{m0}}}
\newcommand\OmLamzero{\Omega_{\Lambda0}} 
 
\newcommand\xiscrad{\xi_{\parallel}^{\mathrm{sc}}}
\newcommand\xisctang{\xi_{\perp}^{\mathrm{sc}}}
\newcommand\xinonscrad{\xi_{\parallel}^{\mathrm{non-sc}}}
\newcommand\xinonsctang{\xi_{\perp}^{\mathrm{non-sc}}}
\newcommand\xivoidrad{\xi_{\parallel}^{\mathrm{void}}}
\newcommand\xivoidtang{\xi_{\perp}^{\mathrm{void}}}
\newcommand\xinonvoidrad{\xi_{\parallel}^{\mathrm{non-void}}}
\newcommand\xinonvoidtang{\xi_{\perp}^{\mathrm{non-void}}}
\newcommand\ralltang{r_{\perp}^0}

\newcommand\rsctang{r_{\perp}^{\mathrm{sc}}}
\newcommand\rnonsctang{r_{\perp}^{\mathrm{non-sc}}}
\newcommand\rvoidtang{r_{\perp}^{\mathrm{void}}}
\newcommand\rnonvoidtang{r_{\perp}^{\mathrm{non-void}}}
\newcommand\DD{\mathrm{DD}}
\newcommand\DR{\mathrm{DR}}
\newcommand\RR{\mathrm{RR}}
\newcommand\Nboot{N_{\mathrm{boot}}}
\newcommand\NRzero{N_{\mathrm{R}}^0}

\newcommand\aeff{a_{\mathrm{eff}}}

\newcommand\superclus{^{\mathrm{SC}}}

\newcommand\bggal{^{\mathrm{D}}}

\newcommand\reffsuperclus{R}

\newcommand\rradialcomov{r}

\providecommand\tablefoot[1]{{#1}}
\providecommand\tablefoottext[1]{${}^{#1}$}

\providecommand\tablefootmarkmath[1]{^{#1}\,\ignorespaces}

\providecommand\filenamestyle[1]{{\tt #1}} 

\begin{document}

\title[Environment-dependent BAO shift]{\protect\postrefereetitle{Evidence for
    an environment-dependent shift in the baryon acoustic oscillation peak}}


\author[Roukema et al.]{Boudewijn
  F. Roukema$^{1,2}$\protect\thanks{{In
      memory of my father, Arie Roukema, who made me curious about the
      fundamental properties of the Universe.}},
  Thomas Buchert$^{2}$,
  Jan J. Ostrowski$^{1,2}$,
  \and
  Martin J. France$^2$
  \\
  $^1$ Toru\'n Centre for Astronomy, 
  Faculty of Physics, Astronomy and Informatics,
  Nicolaus Copernicus University,
  ul. Gagarina 11, 87-100 Toru\'n, 
  Poland
  \\
  $^2$
  Universit\'e de Lyon, Observatoire de Lyon,
  Centre de Recherche Astrophysique de Lyon, CNRS UMR 5574: Universit\'e Lyon~1 and \'Ecole Normale\\ Sup\'erieure de Lyon, 
  9 avenue Charles Andr\'e, F--69230 Saint--Genis--Laval, France\protect\thanks{BFR: during visiting lectureship; 
    JJO: during long-term visit.}}


\date{\frtoday}



\newcommand\Nchainsmain{16}
\newcommand\Npergroup{four}

\maketitle

\begin{abstract}
{\protect\postrefereechanges{The Friedmann--Lema\^{\i}tre--Robertson--Walker (FLRW) metric assumes
  comoving spatial rigidity of metrical
    properties. The curvature term in comoving coordinates 
    is environment-independent and cannot evolve.
    In the standard model,
    structure formation 
    is interpreted accordingly:
    structures average out on the chosen metrical background,
    which remains rigid in comoving coordinates despite 
    nonlinear structure growth.
    The latter claim needs to be tested, since it is 
    a hypothesis that is not derived using general relativity.}}
{We \protect\postrefereechanges{introduce a test of the
    comoving rigidity assumption}
 by measuring the two-point auto-correlation
  function on comoving scales---assuming FLRW comoving spatial rigidity---in order to detect shifts in the 
  \protect\postrefereechanges{baryon} acoustic
  oscillation (BAO) peak location for 
  \protect\postrefereechangesbis{Luminous} Red Galaxy
  (LRG) pairs of the Sloan Digital Sky Survey Data Release 7.
}
{{In} tangential directions, subsets of pairs 
  {overlapping with}
  superclusters or voids show the BAO peak.  The tangential BAO peak
  location 
  {for overlap with}
  Nadathur \& Hotchkiss superclusters is
  $4.3\pm1.6${\hMpc} less than that for LRG pairs unselected for
  supercluster overlap, and $6.6\pm2.8${\hMpc} less than that of the
  complementary pairs.  Liivam{\"a}gi et al.\/ superclusters give
  corresponding differences of $3.7\pm2.9${\hMpc} and
  $6.3\pm2.6${\hMpc}, respectively.  }
{We have found moderately significant evidence (Kolmogorov--Smirnov
  tests suggest very significant evidence) that the BAO peak location
  for supercluster-overlapping pairs is compressed by about 6\%
  compared to that of {the} 
  complementary sample, providing a
  {potential} challenge to FLRW models and a benchmark for predictions from \protect\postrefereechanges{models based on an
    averaging approach that leaves the spatial metric
    {{\em a priori}} unspecified.}}
\end{abstract}


\mystamp{\mystampdothestamp}


\dodouble{ \clearpage } 


\newcommand\tNobjects{
  \begin{table}
    \caption{\mycaptionfont Numbers of SDSS DR7 Northern Galactic Cap LRGs and superclusters
      (\SSS\protect\ref{s-method-catalogues})
      \label{t-Nobjects}}
    $$\begin{array}{c c c c} \hline
      \mathrm{subset}\tablefootmarkmath{a} 
      & D\tablefootmarkmath{b} 
      & R\tablefootmarkmath{c} 
      & \mathrm{ref} 
      \rule{0ex}{2.5ex} 
      \\
      \hline 
      \multicolumn{4}{l}{\mbox{LRGs:}} 
      \rule{0ex}{2.5ex} 
      \\
      \mathrm{dim} &  61899 & 3082871 & \mbox{\protect\citet{Kazin2010}} \\
      \mathrm{bright} & 30272 & 1521736 & \mbox{\protect\citet{Kazin2010}} \\
      \multicolumn{4}{l}{\mbox{superclusters:}} \\
      \mathrm{dim+bright}
      & 235 & {} & \mbox{\protect\citet{NadHot2013}} \\
      z < 0.6\tablefootmarkmath{d} 
      & 2701 & {} & \mbox{\protect\citet{Liivamagi12}} \\
      \multicolumn{4}{l}{\mbox{voids:}} \\
      \mathrm{dim+bright} & 83 & & \mbox{\protect\citet{NadHot2013}} \\
      \hline
    \end{array}$$ \\
    \tablefoot{ 
      \tablefoottext{a}{The ``dim'' and ``bright'' subsets of LRGs are defined by the authors of these analyses.}\\
      \tablefoottext{b}{``Data'': observed galaxies,
        {as listed by 
          \protect\citet{Kazin2010}}.}\\
      \tablefoottext{c}{``Random'': galaxies simulated by \citet{Kazin2010} 
        to mimic selection criteria. Randomly selected subsets of these are
        used in the present work.}\\
      \tablefoottext{d}{See {\SSS}~3.2 of \protect\citet{Liivamagi12} for details.}
    }
  \end{table}
}  

\newcommand\foverlap{
  \begin{figure}
    \centering 
    \includegraphics[width=8cm]{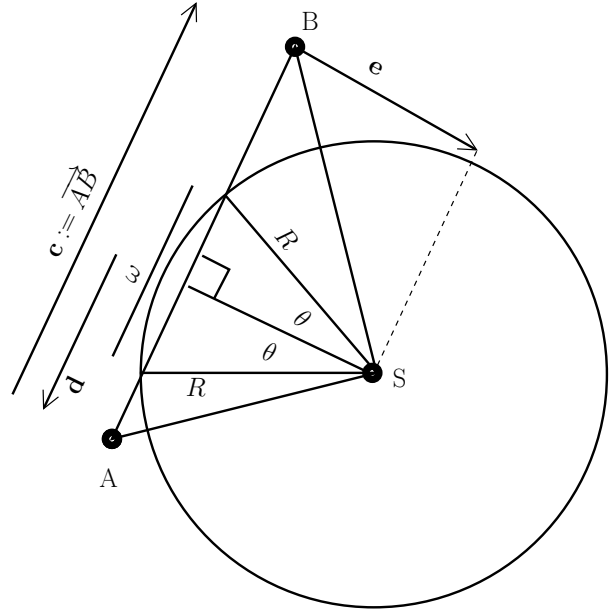}
    \caption[]{ \mycaptionfont Geometrical definition of the overlap
      between a supercluster (or void) and an LRG--LRG pair in the
      comoving observational spatial section interpreted with an FLRW
      flat metric. Two LRGs are located at positions A and B,
      respectively; the supercluster (or void) is centred at S and
      assumed to be spherical, of radius $\reffsuperclus$. The vector
      $\mathbf{e}$ is normal to the pair separation vector
      $\mathbf{c}:=\overrightarrow{\mathrm{AB}}$ 
      {and
      extends from the line segment 
      $\overline{\mathrm{AB}}$ to}
      the supercluster (void)
      centre S.  In the case illustrated, the pair--supercluster overlap
      $\omega$ is the chord length
      [Eq.~\protect\ref{e-overlap-easiest}].  See
      \SSS\protect\ref{s-overlap-defn} for other cases.
    \label{f-overlap}}
  \end{figure} 
} 

\newcommand\fxifull{
  \begin{figure}
    \centering 
    \includegraphics[width=8cm]{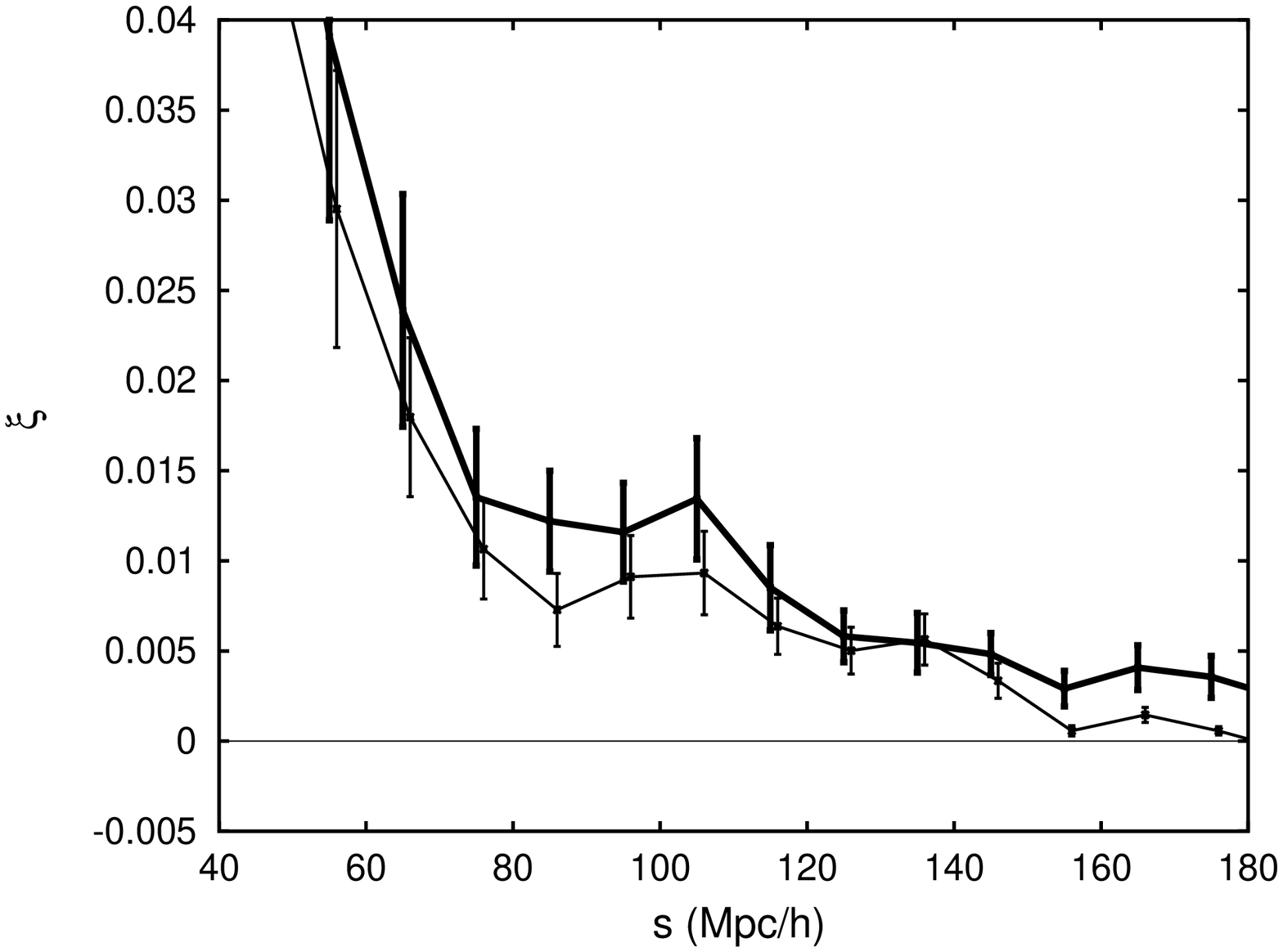}
    \caption[]{ \mycaptionfont Two-point auto-correlation function
      $\xi(s)$ for the ``bright'' (upper curve) and ``dim'' (lower
      curve; horizontally offset by $+1${\hMpc} for clarity) 
      samples
      of luminous red galaxies (LRGs) in the SDSS~DR7, as provided by
      \protect\citet{Kazin2010}, against separation $s$ in ${\hMpc},$
      assuming an effective metric approximated by an FLRW model with
      $(\Ommzero=0.32,\OmLamzero=0.68)$. The error bars show 
      standard deviations of bootstrapped samples
      at each given separation $s$; these error bars
      provide a rough upper limit to the uncertainty 
      and {\em are not used for the 
      analysis}. In each case, the ratio
      $\NRzero/N_{\mathrm{D}}$ is 16 and the 
      number of bootstraps is 16.
      The BAO peak at $\sim 105${\hMpc} is sharp
      for the bright sample and blunt for the dim sample.
      \label{f-xi-full}}
  \end{figure} 
} 

\newcommand\fxiNHsc{
  \begin{figure}
    \centering 
    \includegraphics[width=8cm]{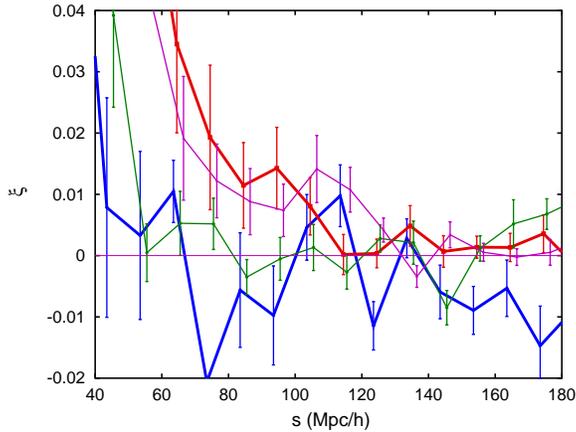}
    \caption[]{ \mycaptionfont Two-point auto-correlation function
      $\xi(s)$ 
      for the SDSS DR7 ``bright'' sample
      assuming an effective metric approximated by an FLRW
      model with $(\Ommzero=0.32,\OmLamzero=0.68)$, subdivided into
      radial (blue, lower, thick curve) and 
      tangential (red, upper, thick curve) components 
      overlapping with 
      \protect\citet{NadHot2013} superclusters, and
      non-overlapping
      radial (green, mostly lower, thin curve) and 
      tangential (purple, mostly upper, thin curve) components.
      For clarity, the four curves are offset horizontally by
      $-1.5, -0.5, 0.5, 1.5${\hMpc}, respectively.
      The ratio
      $\NRzero/N_{\mathrm{D}}=32, \Nboot=8$.
      {The} vertical scale
      differs from that in Fig.~\protect\ref{f-xi-full}.
      As in Fig.~\protect\ref{f-xi-full}, this figure and the
      following show per--separation-bin bootstrap
      standard deviations as error bars, 
      {providing}
      a rough upper limit guide to the uncertainty, 
      but these {\em are not used for statistical
      analysis}.
      \label{f-xi-NH-sc}}
  \end{figure} 
} 

\newcommand\fxiNHvoid{
  \begin{figure}
    \centering 
    \includegraphics[width=8cm]{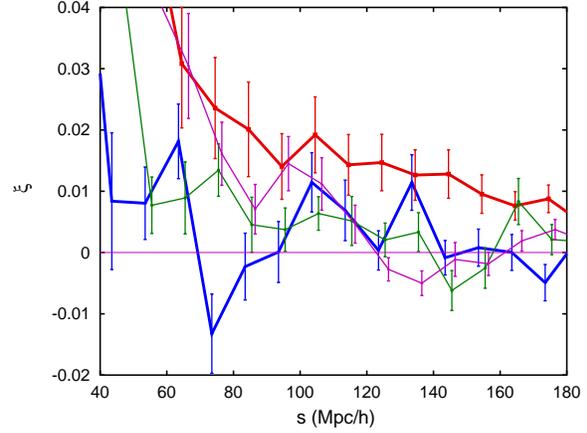}
    \caption[]{ \mycaptionfont 
      As for Fig.~\ref{f-xi-NH-sc}, for voids, i.e.
      $\xi(s)$ for pairs overlapping
      (radial: blue, lower, thick curve; 
      tangential: red, upper, thick curve)
      or not overlapping 
      (radial: green, lower, thin curve;
      tangential: purple, upper, thin curve) 
      with
      \protect\citet{NadHot2013} voids.
      The ratio 
      $\NRzero/N_{\mathrm{D}} =  32, \Nboot =8$.
      \label{f-xi-NH-void}}
  \end{figure} 
} 

\newcommand\fxiLiiva{
  \begin{figure}
    \centering 
    \includegraphics[width=8cm]{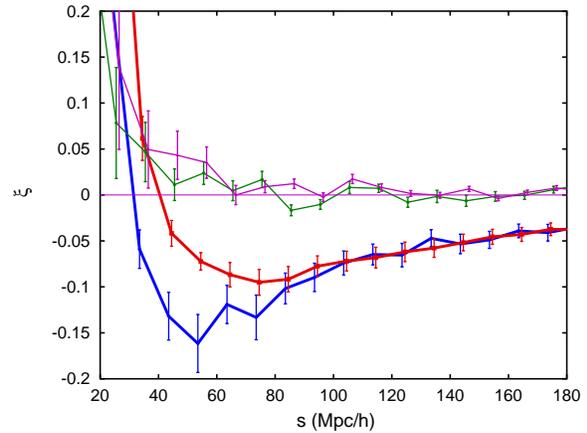}
    \caption[]{ \mycaptionfont As for Fig.~\ref{f-xi-NH-sc}, for
      \protect\citet{Liivamagi12}'s superclusters,
      i.e.
      $\xi(s)$ for pairs overlapping
      (radial: blue, lower, thick curve; 
      tangential: red, upper, thick curve)
      or not overlapping 
      (radial: green, lower, thin curve;
      tangential: purple, upper, thin curve) 
      superclusters.  The vertical scale
      covers a range of about an order of magnitude greater than 
      in the previous plots.  The calculation
      parameters are $\NRzero/N_{\mathrm{D}} = 16, \Nboot=16$.
      The integral constraint (not corrected for) obviously
      has a strong effect for the overlapping components 
      (thick curves).
      \label{f-xi-Liiva}}
  \end{figure} 
} 

\newcommand\fcubicfitfull{
  \begin{figure}
    \centering 
    \includegraphics[width=8cm]{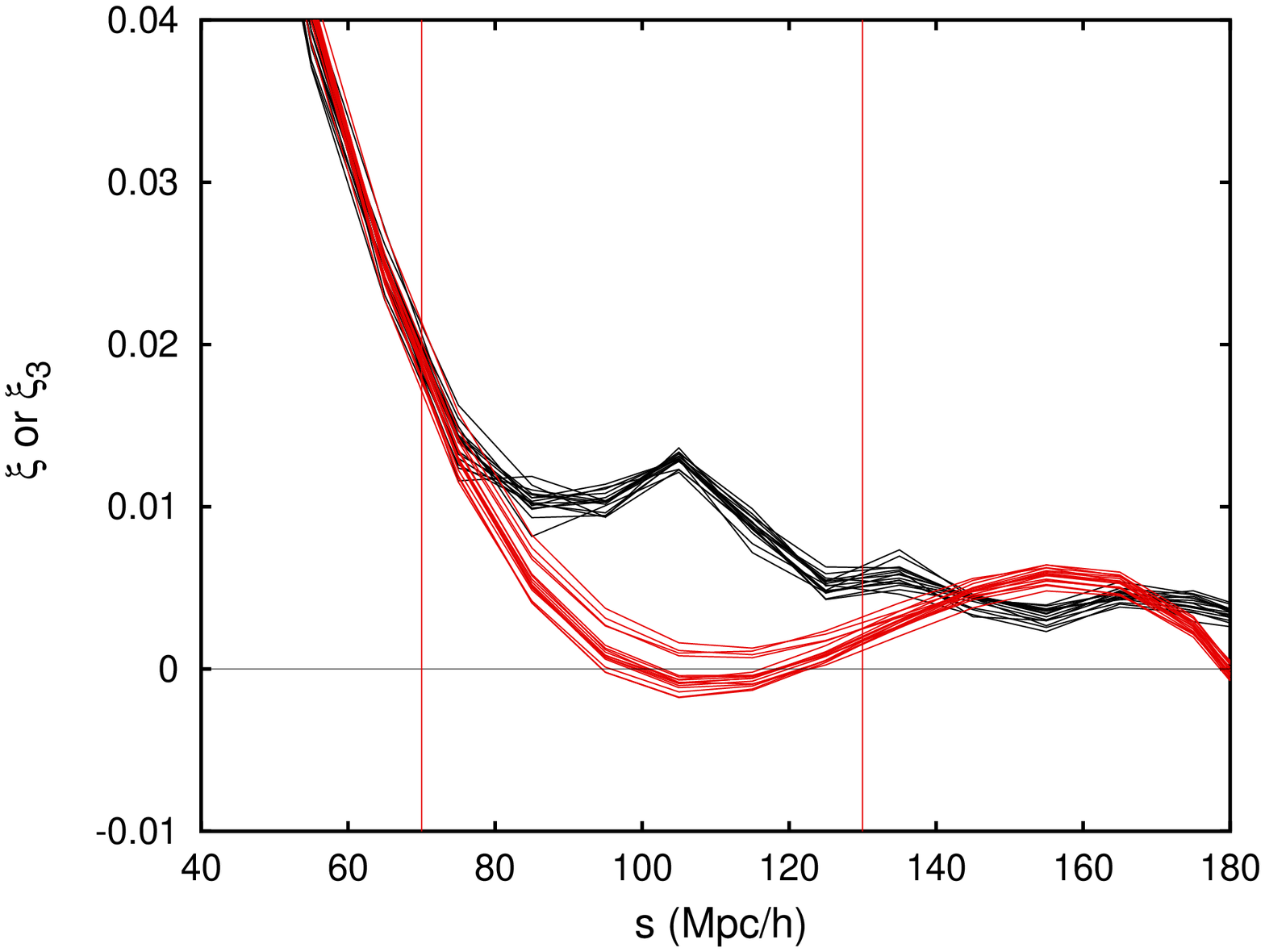}
    \includegraphics[width=8cm]{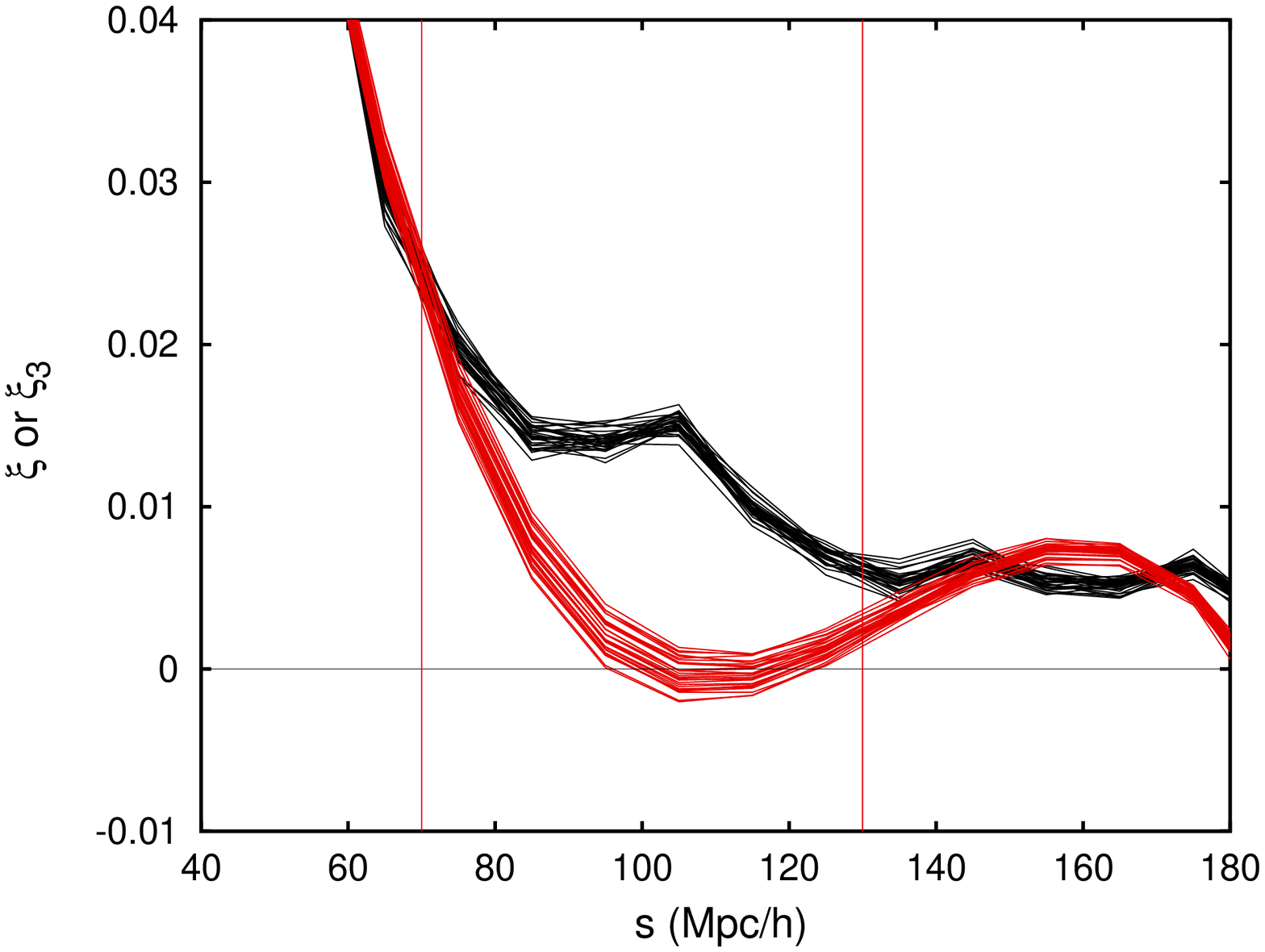}
    \caption[]{ \mycaptionfont      
      {Bootstraps estimates $\xi(s)$ (black online)
      for the full (``bright'')
      LRG sample (with no supercluster or void dependence)
      and best-fit cubics (red online), as described in
      \SSS\protect\ref{s-method-bao-locations}.
      Vertical lines show the range $(s <70) \cup (130 < s)$ 
      (with $s$ in {\hMpc})
      over which the cubics are fit. Subtracting the cubics 
      (Fig.~\ref{f-baopeak-full}) accentuates the BAO peak.
      {\em Upper panel:} pairs in both directions;
      {\em lower panel:} tangential pairs.}
      \label{f-cubicfit-full}}
  \end{figure} 
} 

\newcommand\fbaopeakfull{
  \begin{figure}
    \centering 
    \includegraphics[width=8cm]{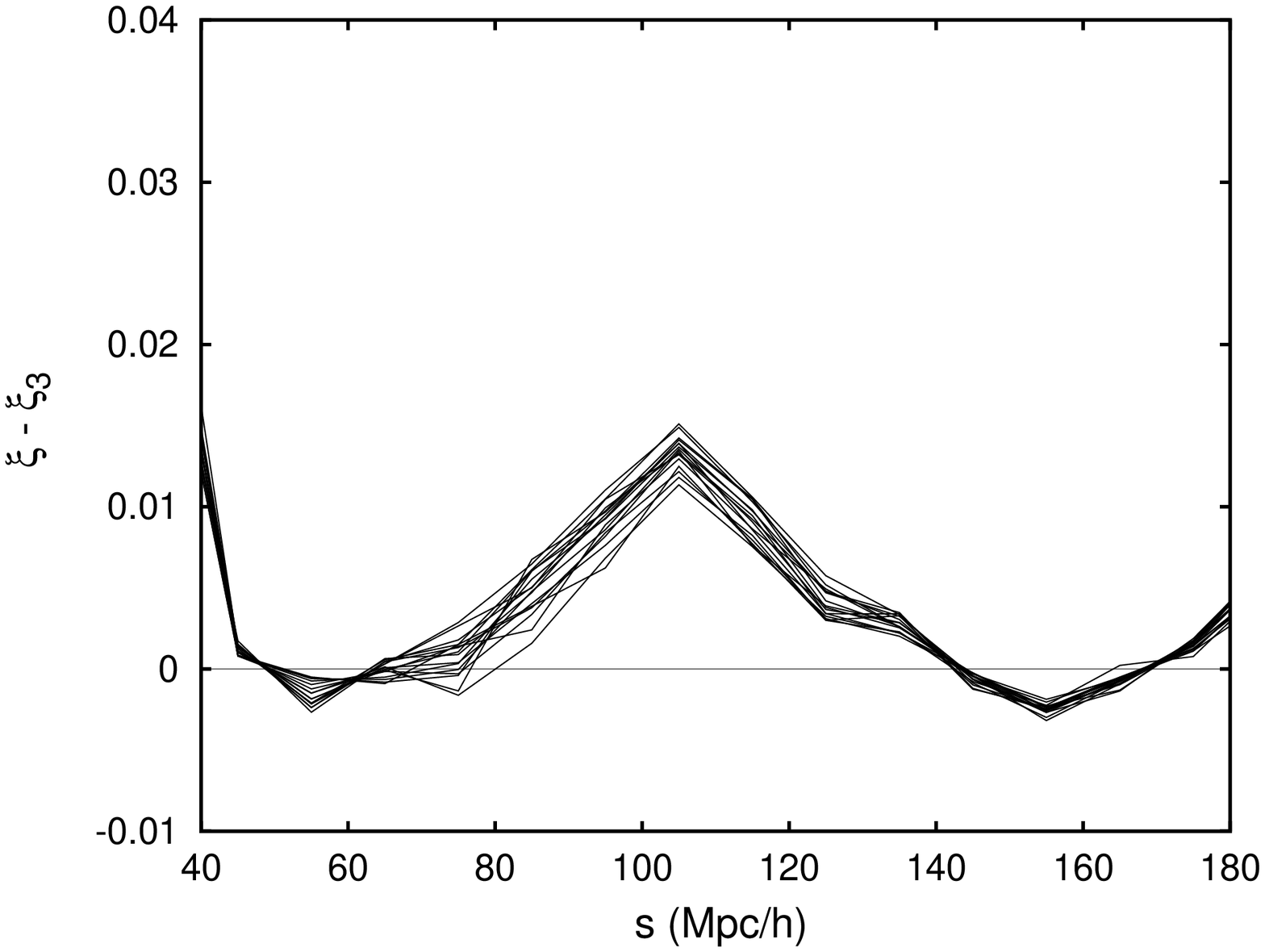}
    \includegraphics[width=8cm]{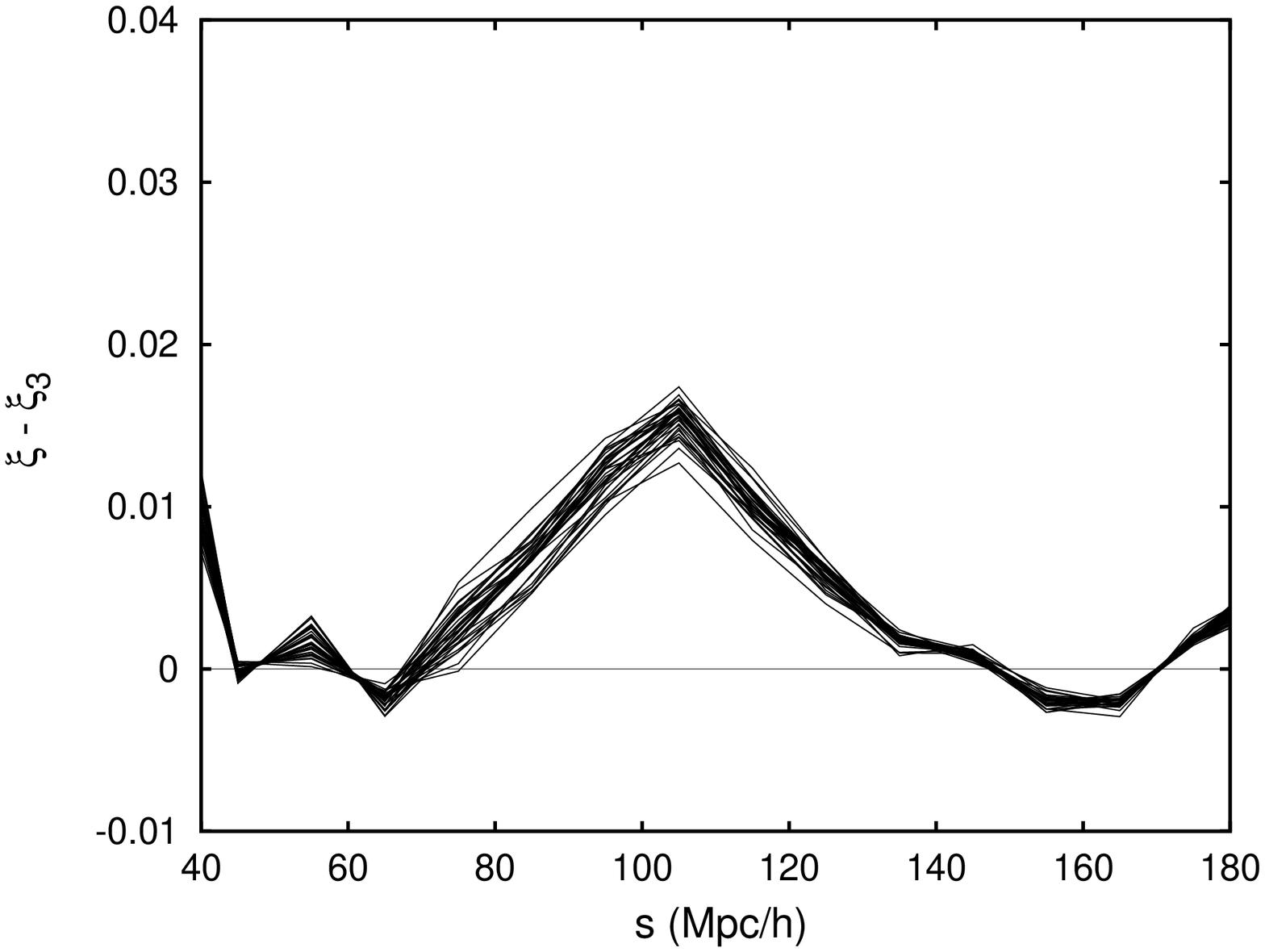}
    \caption[]{ \mycaptionfont Cubic-subtracted
      two-point auto-correlation function
      $\xi(s) - \xi_3(s)$ 
      for the bootstraps of the full (``bright'')
      LRG sample,
      {corresponding to the subtraction of
        the respective curves in Fig.~\protect\ref{f-cubicfit-full}.}
      {\em Upper panel:} pairs in both directions;
      {\em lower panel:} tangential pairs.
      \label{f-baopeak-full}}
  \end{figure} 
} 

\newcommand\fbaopeakNHsc{
  \begin{figure}
    \centering 
    \includegraphics[width=8cm]{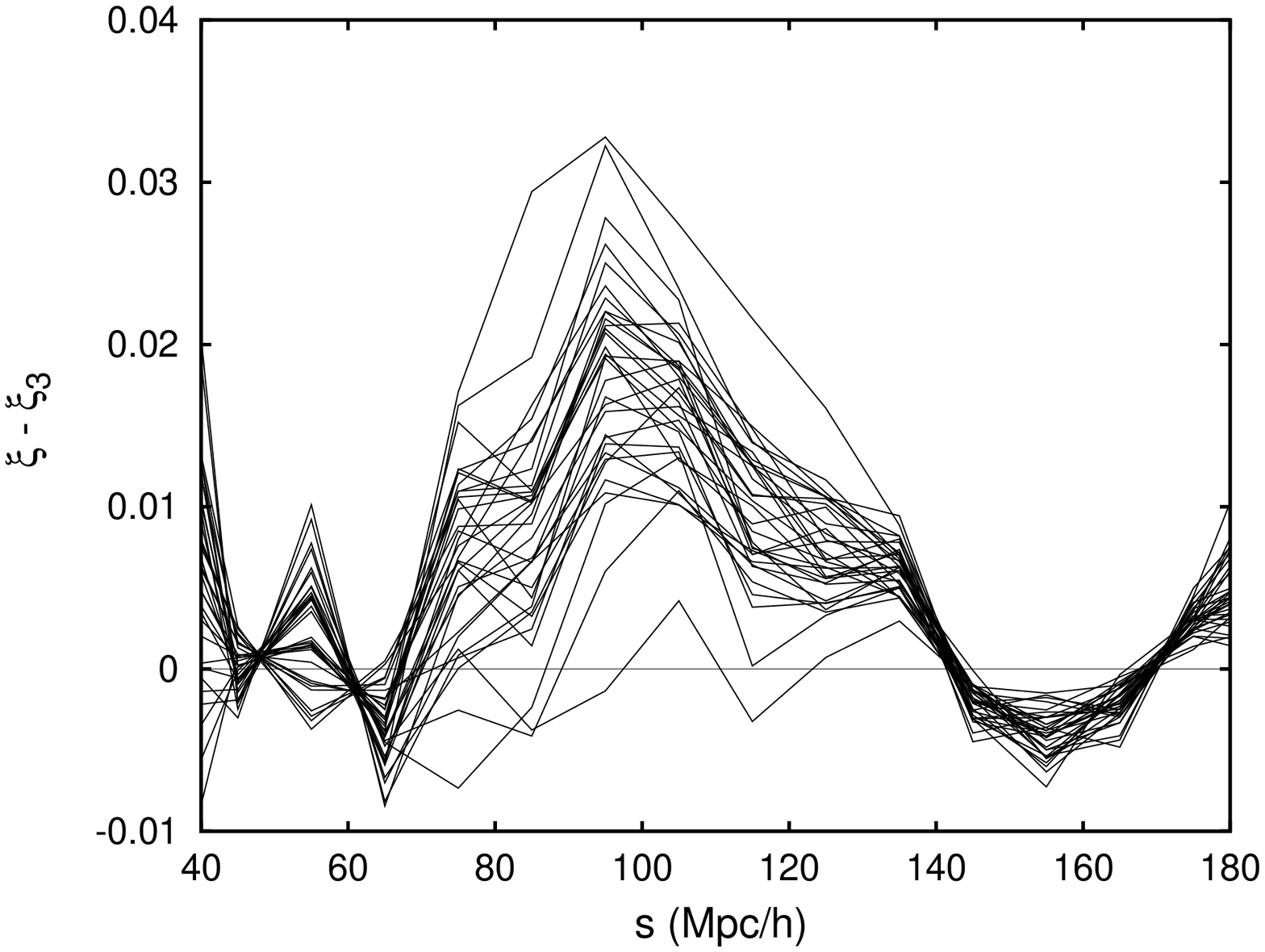}
    \includegraphics[width=8cm]{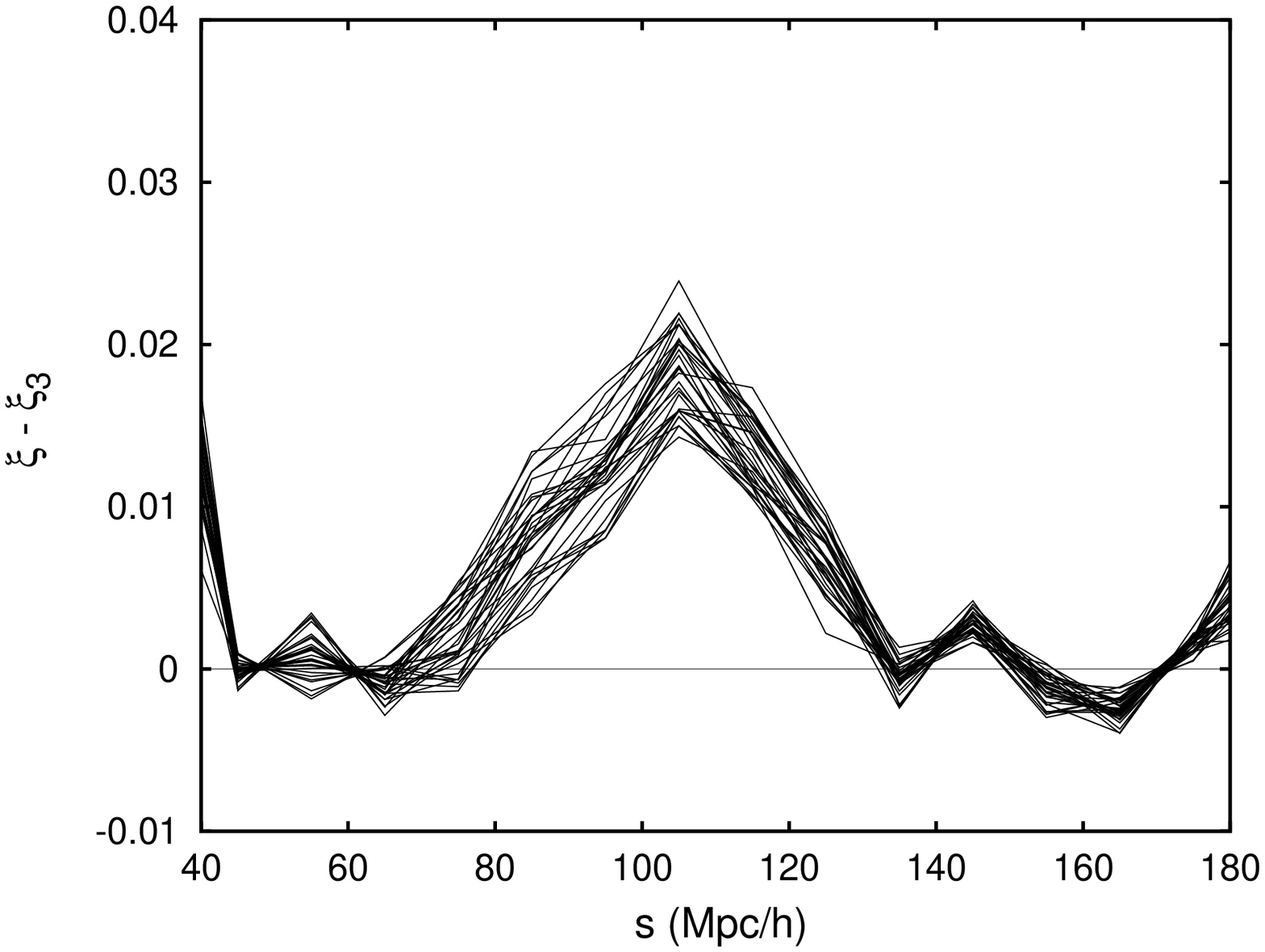}
    \caption[]{ \mycaptionfont Cubic-subtracted
      two-point auto-correlation function
      $\xi(s) - \xi_3(s)$
      for the bootstraps for tangential (``bright'' sample) 
      LRG--LRG pairs 
      near the \protect\cite{NadHot2013} superclusters 
      {\em (above)} and for the complementary set of pairs
      {\em (below)}.
      \label{f-baopeak-NH-sc}}
  \end{figure} 
} 

\newcommand\fcdfNHsc{
  \begin{figure}
    \centering 
    \includegraphics[width=8cm]{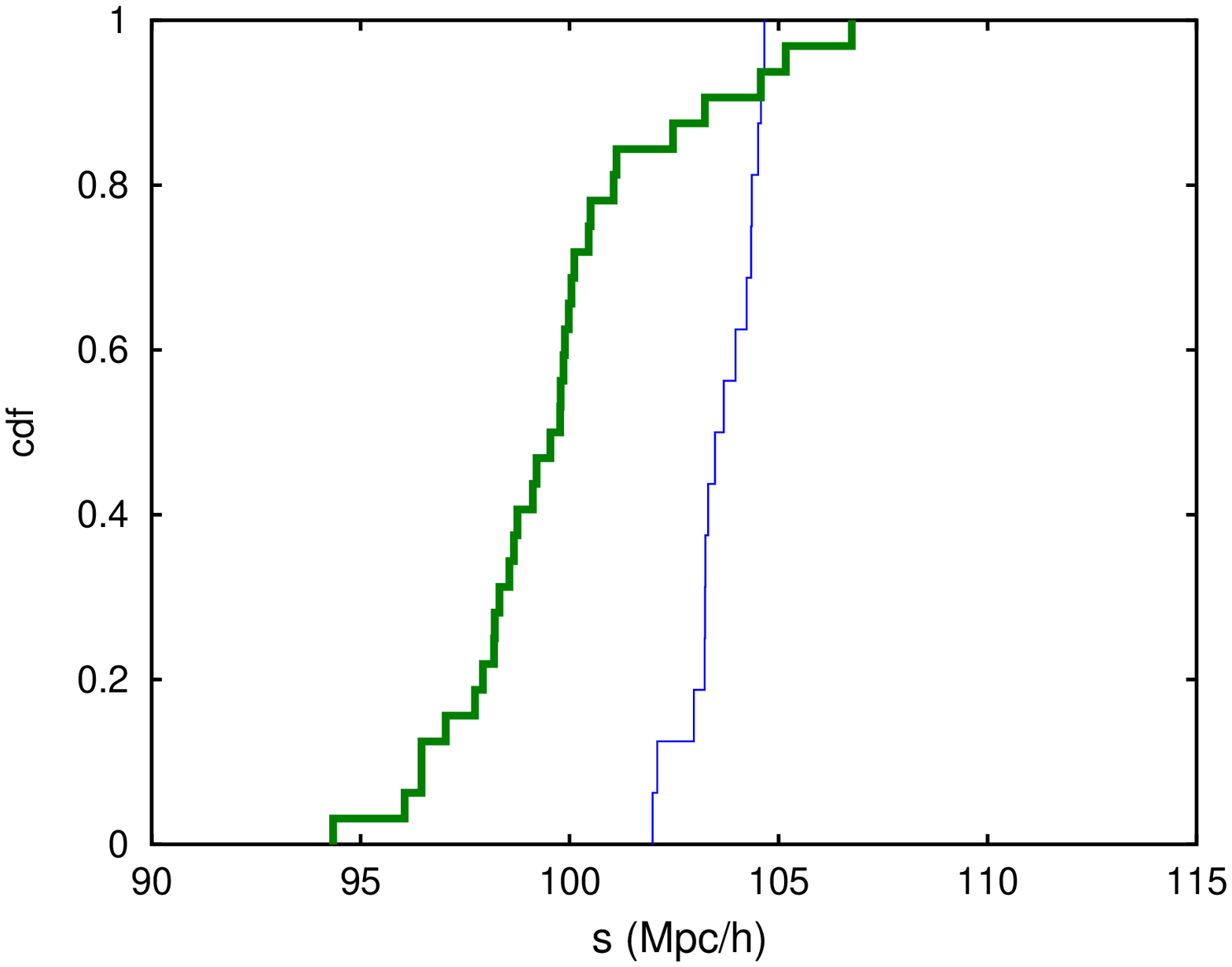}
    \includegraphics[width=8cm]{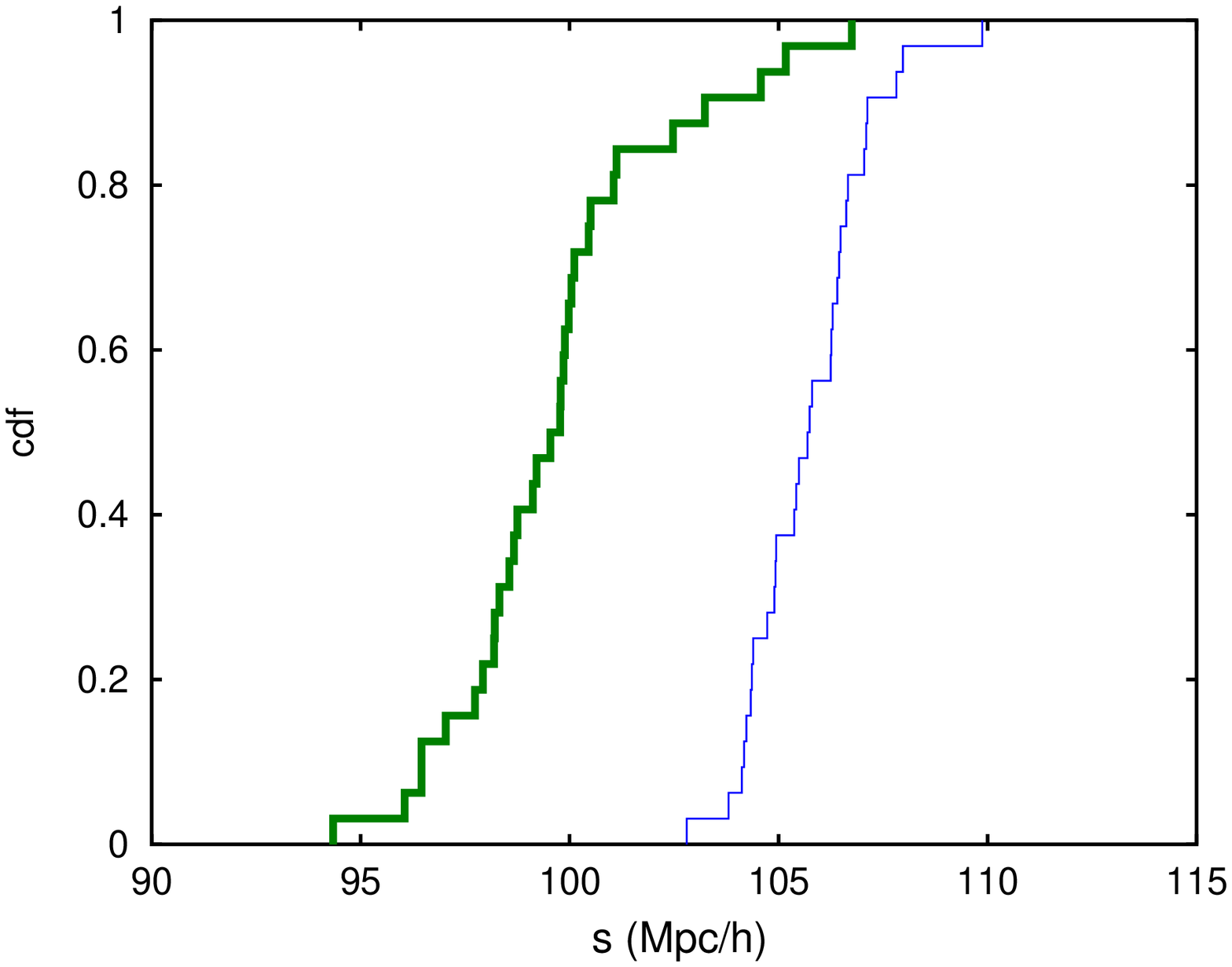}
    \caption[]{ \mycaptionfont Cumulative distribution function
      (cdf) of Gaussian fit BAO peak locations 
      [\SSS\protect\ref{s-method-bao-locations};
        $p_1$ in Eq.~(\ref{e-gaussian-model})] 
      for the bootstraps for tangential (``bright'' sample) 
      LRG--LRG pairs 
      near the \protect\cite{NadHot2013} superclusters,
      shown as a thick (green online) curve in both panels.
      {\em Upper panel:} the cdf for bootstrap BAO peak
      locations for the tangential (``bright'') sample 
      without overlap selection
      is shown as a thin (blue
      online) curve; {\em lower panel:} the cdf for
      the complementary set of
      pairs (those not overlapping with superclusters)
      is shown as a thin (blue
      online) curve.
      The lower panel corresponds to the two panels of
      Fig.~\protect\ref{f-baopeak-NH-sc}.
      \label{f-cdf-NH-sc}}
  \end{figure} 
} 

\newcommand\fbaopeakNHvoid{
  \begin{figure}
    \centering 
    \includegraphics[width=8cm]{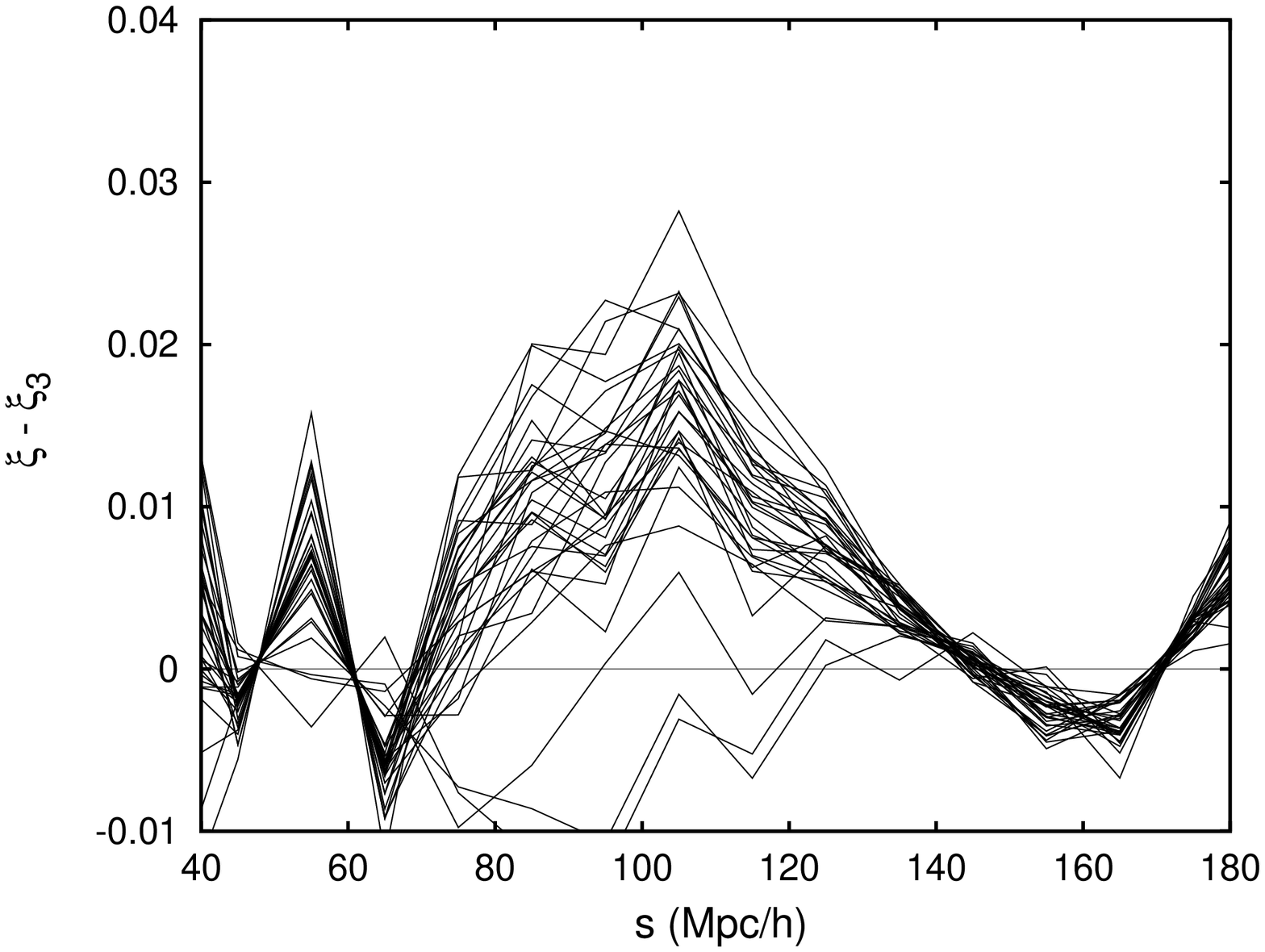}
    \includegraphics[width=8cm]{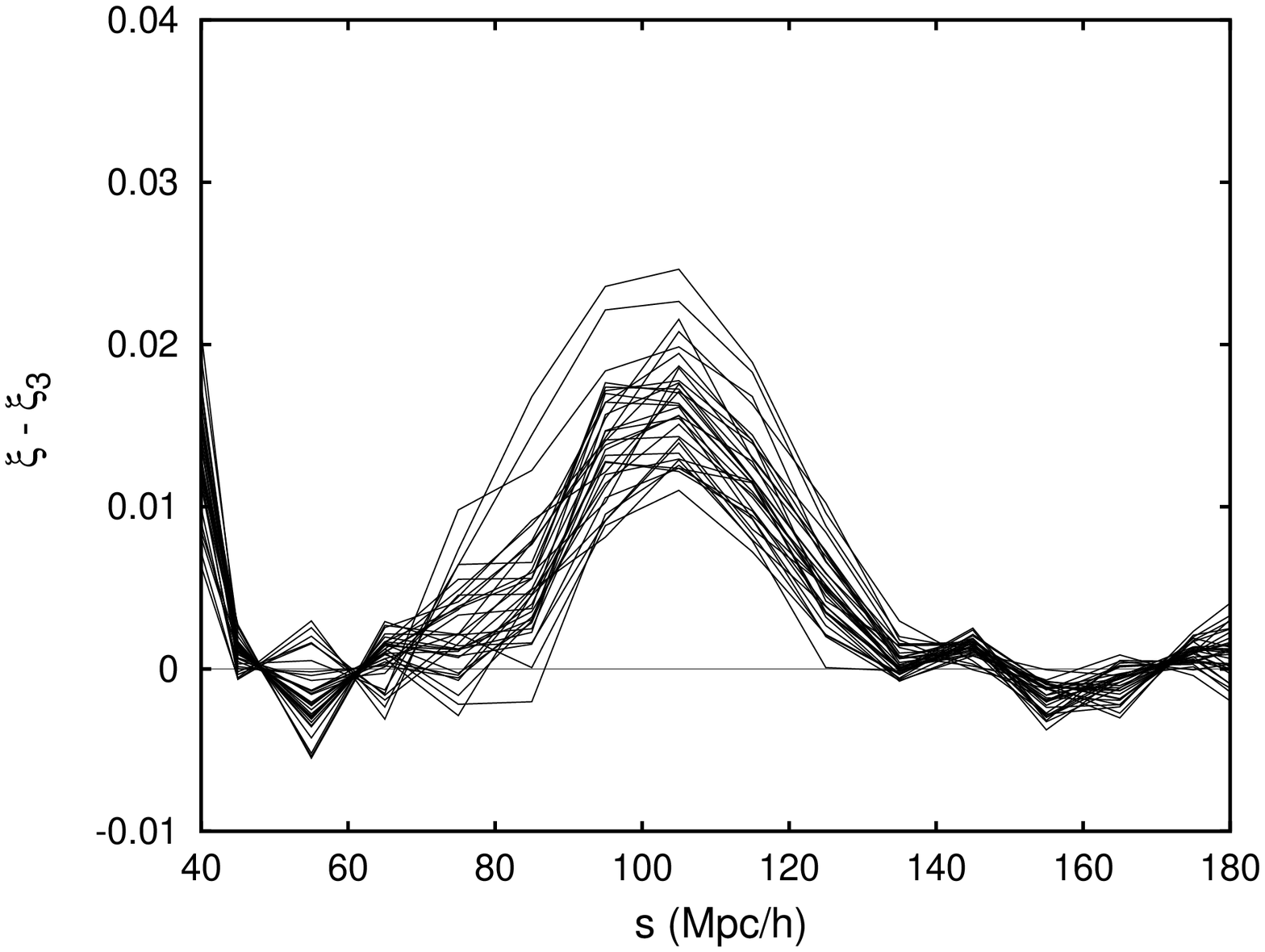}
    \caption[]{ \mycaptionfont Cubic-subtracted
      two-point auto-correlation function
      $\xi(s) - \xi_3(s)$
      for the bootstraps for tangential (``bright'' sample) 
      LRG--LRG pairs 
      near the \protect\cite{NadHot2013} voids
      {\em (above)} and for the complementary set of pairs
      {\em (below)}.
      \label{f-baopeak-NH-void}}
  \end{figure} 
} 

\newcommand\fcdfNHvoid{
  \begin{figure}
    \centering 
    \includegraphics[width=8cm]{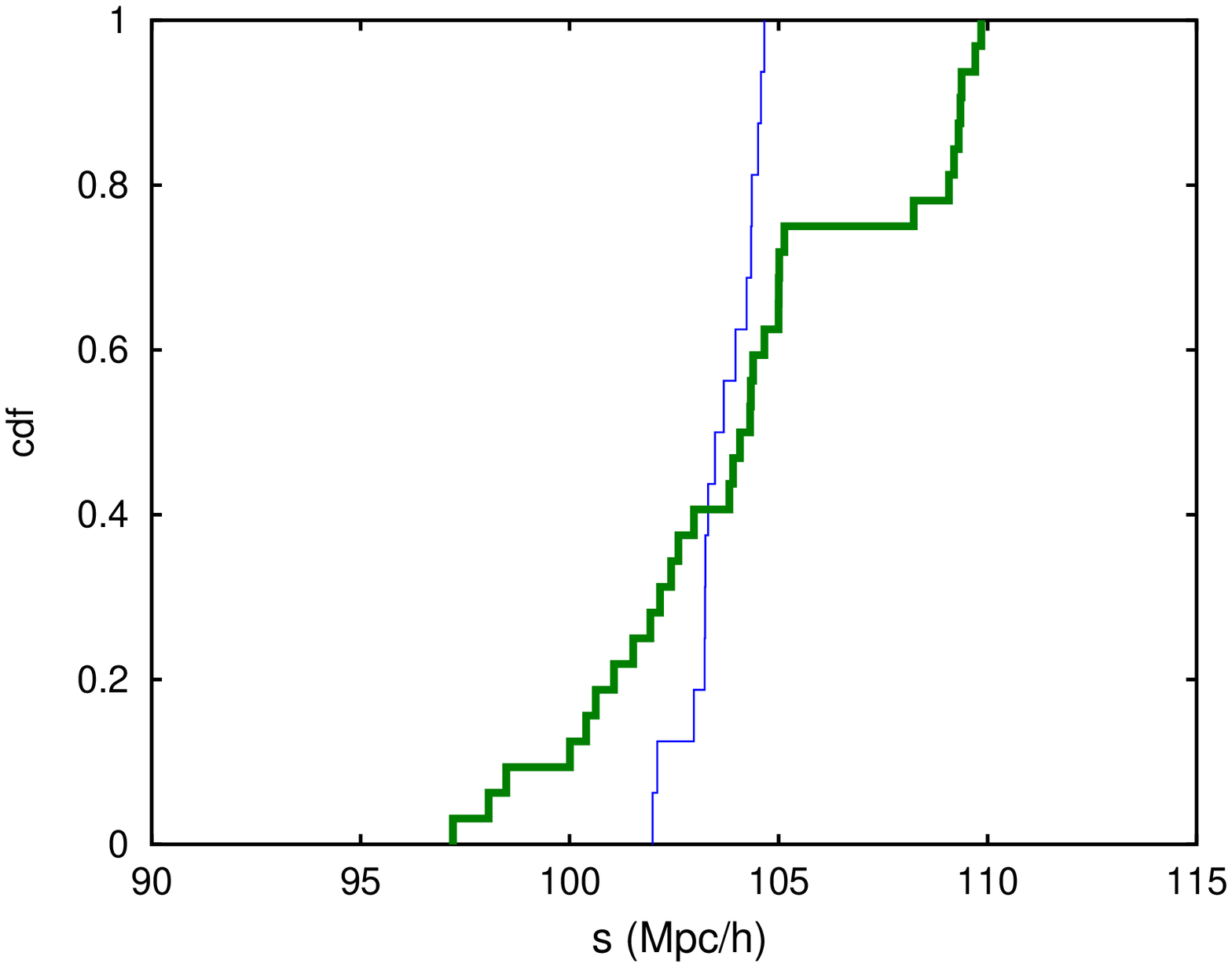}
    \includegraphics[width=8cm]{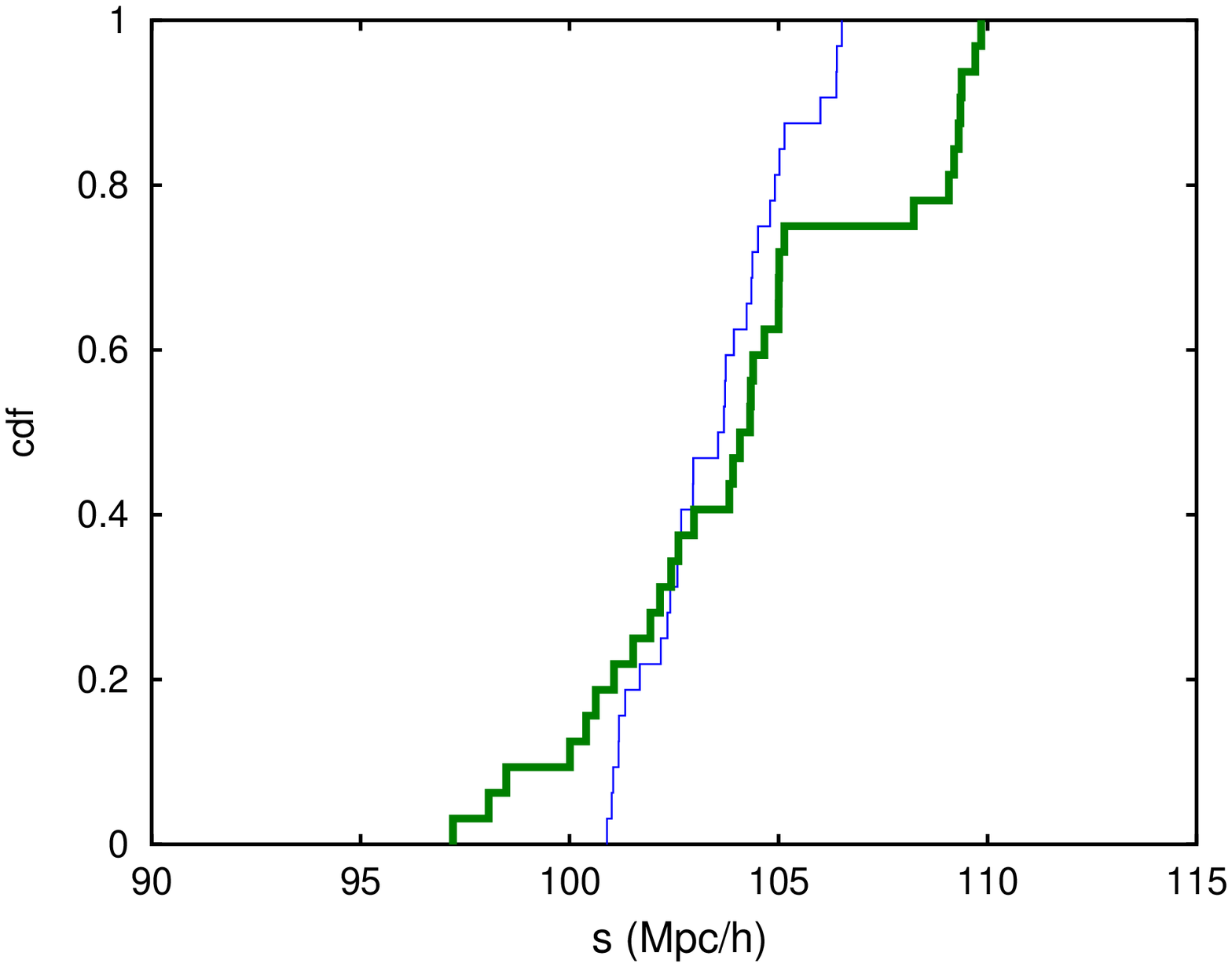}
    \caption[]{ \mycaptionfont As for Fig.~\protect\ref{f-cdf-NH-sc},
      cdf of BAO peak locations for 
      tangential LRG--LRG pairs 
      near the \protect\cite{NadHot2013} voids,
      shown as a thick (green online) curve in both plots,
      compared to that for the tangential sample without
      selection
      (thin curve, blue online, upper plot) and to that
      for pairs that do not overlap with voids
      (thin curve, blue online, lower plot).
      The lower panel corresponds to the two panels of
      Fig.~\protect\ref{f-baopeak-NH-void}.
      \label{f-cdf-NH-void}}
  \end{figure} 
} 

\newcommand\fbaopeakLiiva{
  \begin{figure}
    \centering 
    \includegraphics[width=8cm]{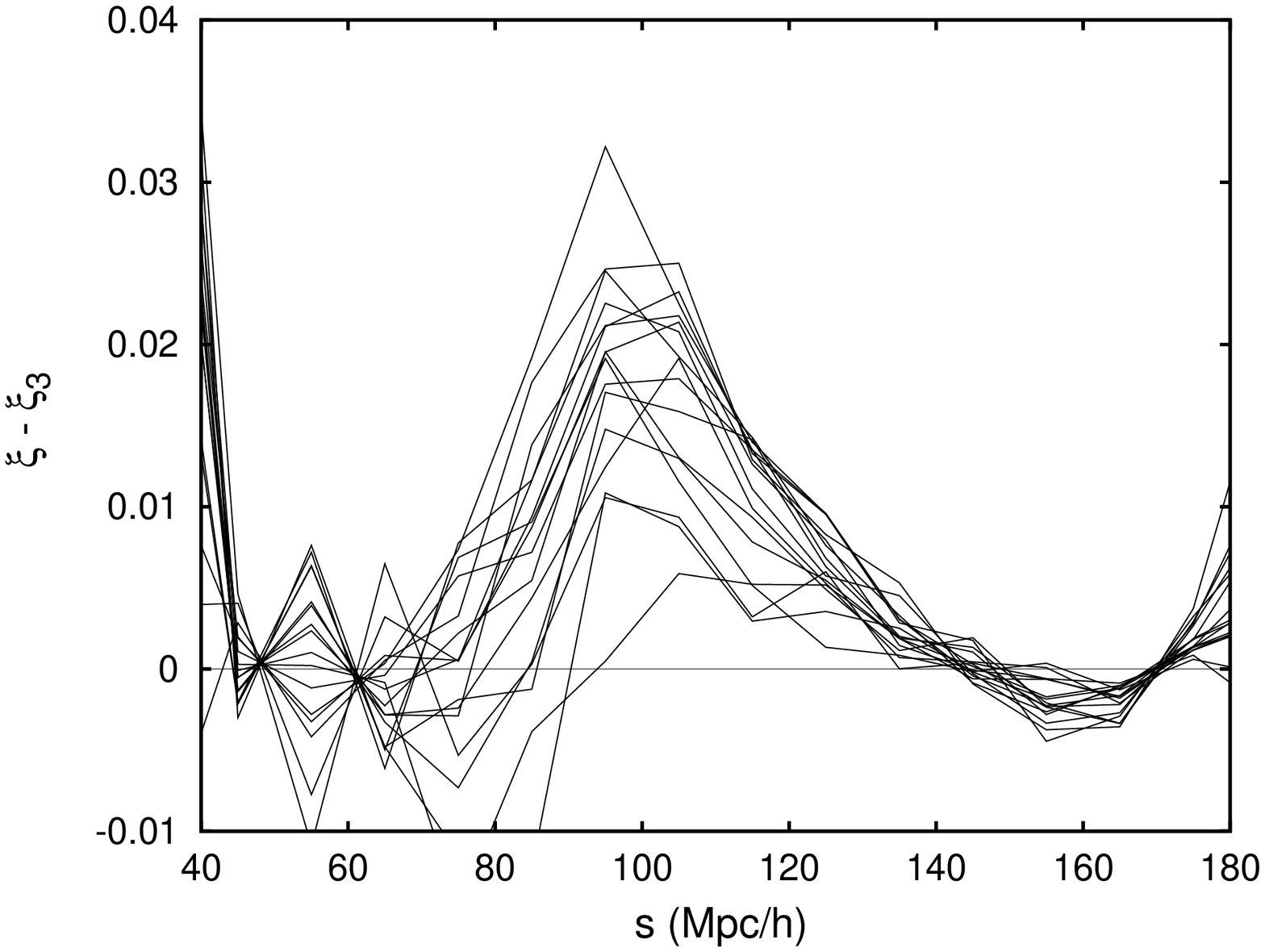}
    \includegraphics[width=8cm]{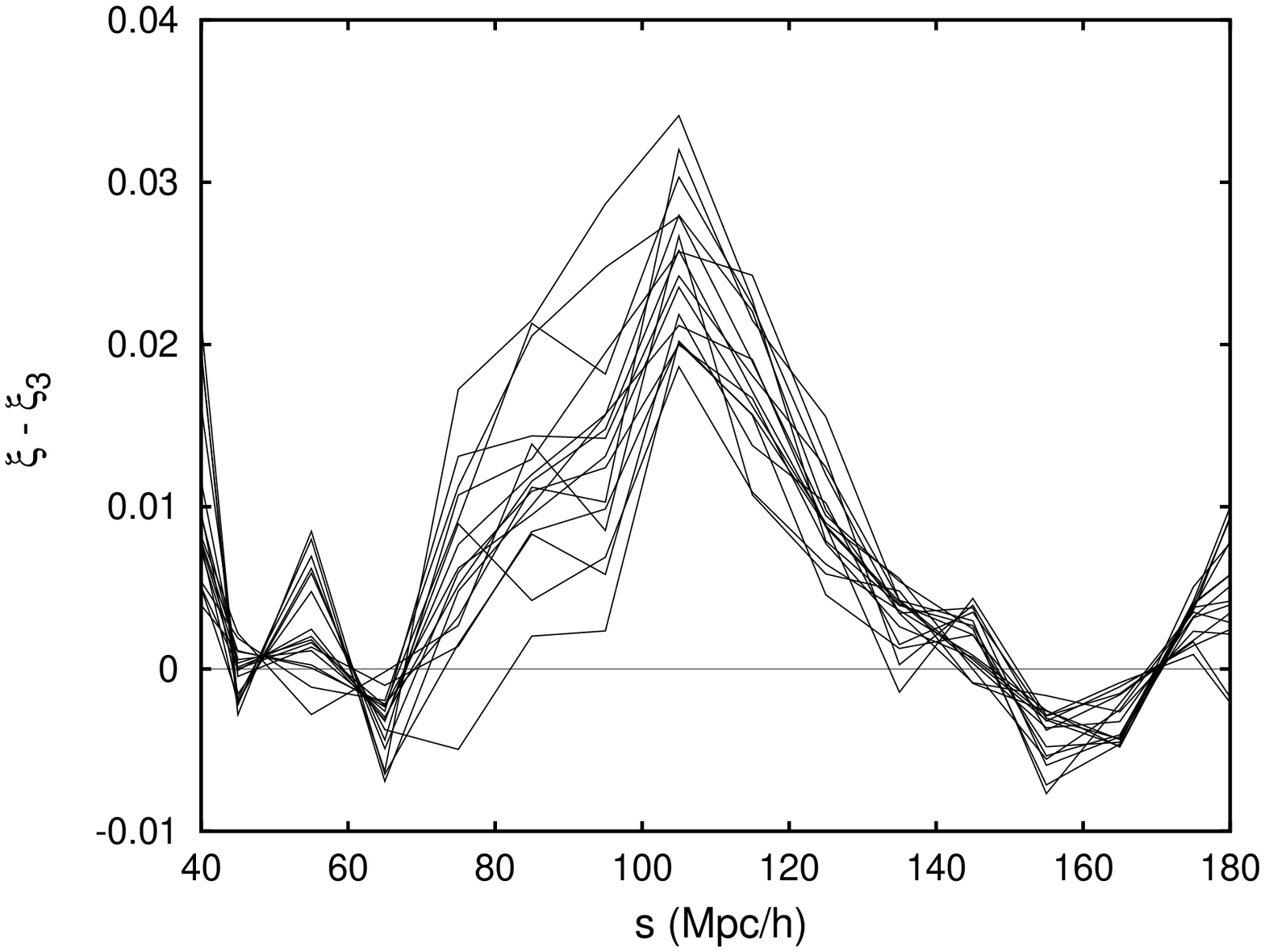}
    \caption[]{ \mycaptionfont Cubic-subtracted
      two-point auto-correlation function
      $\xi(s) - \xi_3(s)$
      for the bootstraps for tangential (``bright'' sample) 
      LRG--LRG pairs 
      near the \protect\cite{Liivamagi12} superclusters 
      {\em (above)} and for the complementary set of pairs
      {\em (below)}.
      {Since this shows the cubic-subtracted correlation function,
      the vertical scale is identical to that of
      Figs~\protect\ref{f-baopeak-full}, \protect\ref{f-baopeak-NH-sc} and \protect\ref{f-baopeak-NH-void},
      in contrast to 
      Fig.~\protect\ref{f-xi-Liiva}, which has a different vertical scale to those of
      Figs~\protect\ref{f-xi-full}, \protect\ref{f-xi-NH-sc} and \protect\ref{f-xi-NH-void}.
      In comparison to Fig.~\protect\ref{f-xi-Liiva}, per-bin variance is reduced, 
      because cubic-subtraction removes much of the noise introduced by the per-bootstrap
      integral constraint.}
      \label{f-baopeak-Liiva}}
  \end{figure} 
} 

\newcommand\fcdfLiiva{
  \begin{figure}
    \centering 
    \includegraphics[width=8cm]{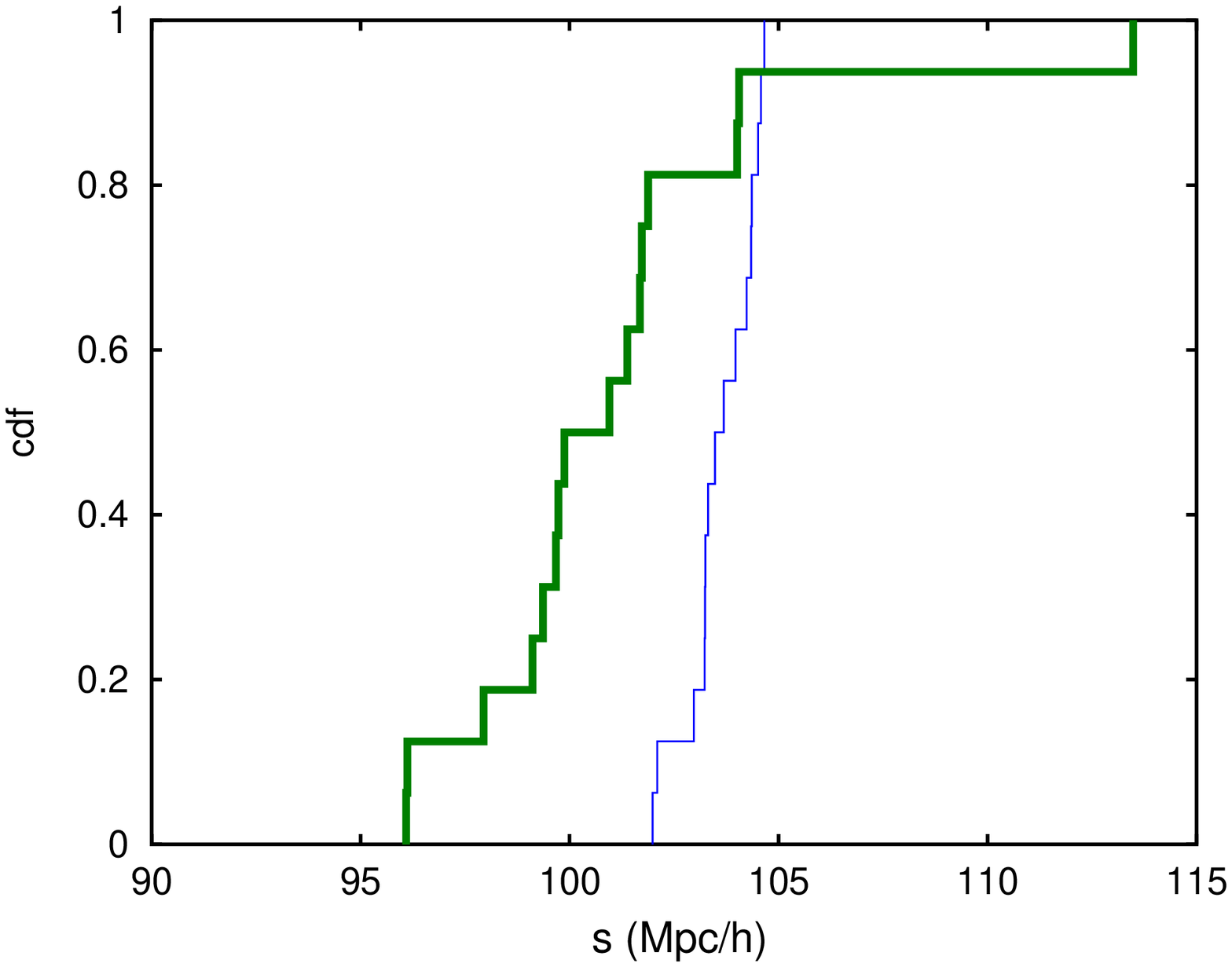}
    \includegraphics[width=8cm]{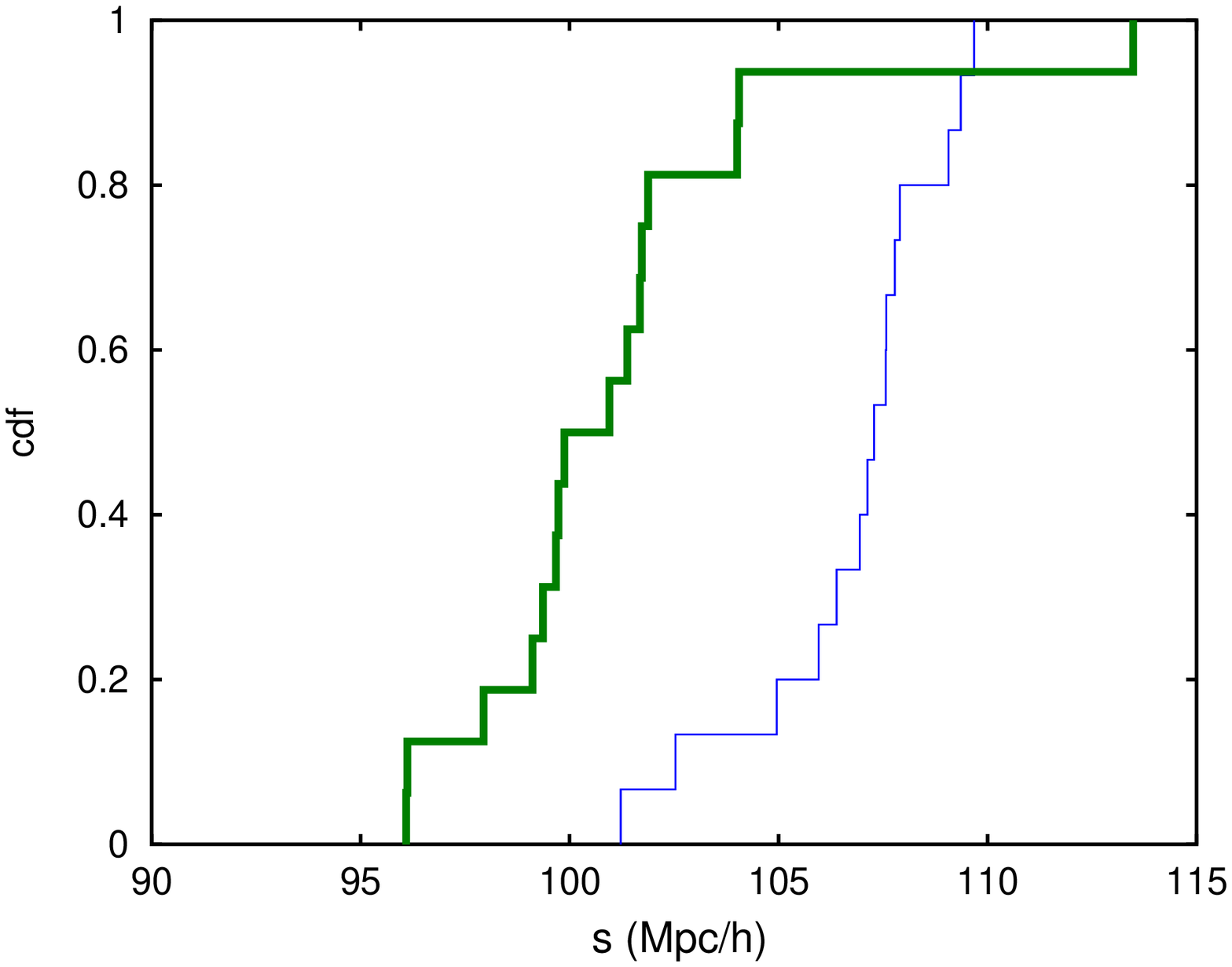}
    \caption[]{ \mycaptionfont As for Fig.~\protect\ref{f-cdf-NH-sc},
      cdf of BAO peak locations for 
      tangential LRG--LRG pairs 
      near the \protect\cite{Liivamagi12} superclusters,
      shown as a thick (green online) curve in both plots,
      compared to that for the tangential full sample 
      (thin curve, blue online, upper plot) and to that
      for pairs that do not overlap with superclusters
      (thin curve, blue online, lower plot).
      The lower panel corresponds to the two panels of
      Fig.~\protect\ref{f-baopeak-Liiva}.
      \label{f-cdf-Liiva}}
  \end{figure} 
} 

\newcommand\tkolsmirnov{
  \begin{table}
    \caption{\mycaptionfont 
      Kolmogorov--Smirnov probabilities of identical 
      cumulative distribution functions
      \label{t-kol-smirnov}}
    $$\begin{array}{c c c c c} 
      \hline 
      \mathrm{catalogue} &
      \ralltang , \rsctang &
      \rnonsctang , \rsctang &
      \ralltang , \rvoidtang &
      \rnonvoidtang , \rvoidtang 
      \rule[-1.5ex]{0ex}{4ex} 
      \\
      \hline
      {\mbox{N\&H}\tablefootmarkmath{a}} 
      \rule{0ex}{2.8ex} 
      &
      3\times10^{-10} & 5\times10^{-11} 
      \,{}\tablefootmarkmath{b}
      &
      0.05 & 0.3 
      \,\tablefootmarkmath{c}
      \\
      {\mbox{LTS}\tablefootmarkmath{d}} &
      5\times10^{-5} & 3\times10^{-5}  
      \,{}\tablefootmarkmath{e}
      \\
      \hline
    \end{array}$$ \\
    \tablefoot{ 
      \tablefoottext{a}{\protect\citet{NadHot2013}.}\\
      \tablefoottext{b}{Columns 1, 2 correspond to
        Fig.~\protect\ref{f-cdf-NH-sc}.}\\
      \tablefoottext{c}{Columns 3, 4 correspond to
        Fig.~\protect\ref{f-cdf-NH-void}.}\\
      \tablefoottext{d}{\protect\citet{Liivamagi12}.}\\
      \tablefoottext{e}{Columns 1, 2 correspond to
        Fig.~\protect\ref{f-cdf-Liiva}.}\\
    }
  \end{table}
} 

\newcommand\tcompression{
  \begin{table}
    \caption{\mycaptionfont Compression${}\tablefootmarkmath{a}$ of BAO peak location
      in {\hMpc}
      \label{t-compression}}
    $$\begin{array}{c c c c c} 
      \hline 
      \mathrm{catalogue} &
      \ralltang - \rsctang &
      \rnonsctang - \rsctang &
      \ralltang - \rvoidtang &
      \rnonvoidtang - \rvoidtang 
      \rule[-1.5ex]{0ex}{4ex} 
      \\
      \hline
      \rule{0ex}{2.8ex} 
      \mbox{N\&H}\tablefootmarkmath{b} &
      4.3\pm1.6 & 6.6\pm2.8 
      \tablefootmarkmath{c}  & 
      -0.2\pm4.0 & -1.1\pm5.5 
      \tablefootmarkmath{d}\tablefootmarkmath{e} \\
      \mbox{LTS}\tablefootmarkmath{f} &
      3.7\pm2.9 & 6.3\pm2.6
      \tablefootmarkmath{g}\tablefootmarkmath{h}  \\ 
      \hline
    \end{array}$$ \\
    \tablefoot{ 
      \tablefoottext{a}{A positive (negative) value in
        columns 1 or 3 indicates compression (stretching) compared to
        the full sample BAO peak location.  A positive (negative)
        value in columns 2 (4) indicates compression (stretching) of
        the BAO peak location for overdensity-selected
        (underdensity-selected) pairs compared to the complementary
        set of
        pairs.}\\
      \tablefoottext{b}{\protect\citet{NadHot2013}.}\\
      \tablefoottext{c}{Columns 1, 2: $\NRzero/N_{\mathrm{D}}=32$, $\Nboot = 32$.}\\
      \tablefoottext{d}{Columns 3, 4: $\NRzero/N_{\mathrm{D}}=32$, $\Nboot = 32$.}\\
      \tablefoottext{e}{Columns 3, 4: The Gaussian BAO peak fit failed
        for one of the 32 bootstraps and was thus ignored in the
        estimate of $\rvoidtang$.}\\
      \tablefoottext{f}{\protect\citet{Liivamagi12}.}\\
      \tablefoottext{g}{Columns 1, 2: $\NRzero/N_{\mathrm{D}}=16$, $\Nboot = 16$.}\\
      \tablefoottext{h}{Columns 1, 2: Gaussian BAO peak fits failed
        for two of the 16 bootstraps and were thus ignored in the
        estimate of $\rsctang$ for the \protect\citet{Liivamagi12} 
        superclusters.}\\
    }
  \end{table}
}  

\newcommand\tnpairfractions{
  \begin{table}
    \caption{\mycaptionfont Fraction of LRG--LRG pairs that overlap
      with superclusters or voids${}\tablefootmarkmath{a}$
      \label{t-npair-fractions}}
    $$\begin{array}{l c c c} 
      \hline 
      \rule[-1.5ex]{0ex}{4ex} 
      \mbox{catalogue} &
      f_{\DD} & f_{\DR}\tablefootmarkmath{b} 
      & f_{\RR}\tablefootmarkmath{b} 
      \\
      \hline
      \rule{0ex}{2.8ex} 
      \mbox{N\&H superclusters}\tablefootmarkmath{c} &
      0.79\pm0.24  &0.76\pm0.23  &0.80\pm0.02 \\
      \mbox{N\&H voids}\tablefootmarkmath{c} &
      0.82\pm0.19  &0.60\pm0.19  &0.83\pm0.07 \\
      \mbox{LTS superclusters}\tablefootmarkmath{d} &
      0.95\pm0.01  &0.72\pm0.02  &0.91\pm0.01  \\
      \hline
    \end{array}$$
    \tablefoot{
      \tablefoottext{a}{Median and 
        $1.4826$ times the median absolute deviation 
        \protect\citep{Hampel74} of the overlapping 
        fraction for the bootstrap realisations,
        for the three types of pairs.}\\
      \tablefoottext{b}{This table is calculated for
        $\NRzero/N_{\mathrm{D}}=1$;
        and $\Nboot = 32, 32, 16,$ respectively, for
        the three cases.}\\
      \tablefoottext{c}{\protect\citet{NadHot2013}.}\\
      \tablefoottext{d}{\protect\citet{Liivamagi12}.}\\
    }
  \end{table}
} 


\section{Introduction}  \label{s-intro}

The low-redshift ($z \ll 3$) Universe has an inhomogeneous density
{distribution} \citep[e.g.][]{deLappGH86},
i.e. during the epoch when most virialisation takes place.  The
standard cosmology model assumes that the nonlinear structure growth
of the virialisation epoch takes place in a spacetime with the
Friedmann--Lema\^{\i}tre--Robertson--Walker (FLRW) metric, and that this
metric is unaffected by structure growth, despite the Einstein
equation.  In other words, an implicit assumption of the FLRW models,
including the $\Lambda$CDM model
\citep[e.g.][]{CosConcord95,WMAPSpergel,PlanckXVIcosmoparam13}, is that
comoving space is rigid, 
a property that is shared by Newtonian models
(see \SSS~2.5 in \citealt{Buch00Hiroshima}, and
\citep{RoukGRF13}).
\postrefereechanges{This assumption needs to be tested.
We propose to test the rigidity of 
a well-established comoving-scale standard ruler,
by examining whether it is environment-dependent.}

{Motivated by
  \citet{EllisStoeger87},
the} scalar averaging approach to cosmology 
{(\citealt*{BKS00,
    Buch00Hiroshima,Buch01scalav};
  \citealt{KolbMNR05};
  \citealt*{Kolb05a,
    Rasanen06superhoriz,
    Rasanen06GRF,
    BuchLarAl06morph,
    Wiltshire07clocks,
    Wiltshire07exact,
    Buchert08status,
    BuchCarf08,
    Wiltshire09timescape,
    Kolb11FOCUS,
    BuchRZA2,
    DuleyWilt13})}
is a general-relativistic
approach to cosmology that extends the Friedmann and acceleration 
equations to the case of general inhomogeneous
distributions of matter {\em and geometry}. 
Initial observational tests show promising
results for template metric implementations of the approach
and for related toy models
(e.g. \citealt{Larena09template};
{\citealt*{ROB13};}
\citealt{BoehmRasan13,Chiesa14Larenatest}; see also models 
with the inclusion
of a phenomenological lapse function: 
\citealt{SmaleWilt11SNe,Wiltshire12Hflow}
or using a different averaging approach: 
\citealt{CCCS12Zalalet}).

Comoving space in {these approaches} is not rigid. This implies that
comoving standard rulers should be inferred to be variable in comoving length
if data are interpreted according to a phenomenological, rigid model, e.g. a
$\Lambda$CDM model. It is generally expected 
\citep{HosoyaBuchMor04,
  Rasanen06superhoriz,
  BuchCarf08,
  ROB13}
that voids should be hyperbolic and that superclusters should 
occupy positively curved
space. 
This is a kinematical expectation, whose details depend on 
whether we interpret the backreaction 
variables in a background-free way,
or whether we use template metrics to 
interpret those variables \citep{BuchCarf03}.
When using a template metric, or a best-fit FLRW metric,
voids and superclusters and their nearby surroundings should
approximately correspond 
to negatively curved, fast expanding regions and
to positively curved regions approaching their turnaround epochs,
respectively.

A comoving standard ruler at $L \sim 100${\hMpc}, the 
\postrefereechanges{baryon} acoustic oscillation (BAO)
peak, has been detected
to high statistical significance \citep[and references thereof]{Eisenstein05MNRAS}, where the
comoving length is determined from the theory of \postrefereechanges{baryon} acoustic
{oscillations \citep[e.g.][and references therein]{EisenHu98Tofk}.
Measuring the BAO peak location---an imprinted feature in comoving space---through either large voids or
superclusters for a given catalogue 
interpreted in rigid (FLRW) comoving space} 
should show
either a stretched or compressed BAO peak location, respectively, in
comparison to 
the BAO peak location for the full catalogue,
\postrefereechanges{interpreted} in the same way.
{However, since} voids dominate spatial 3-volume at the present epoch, 
the scalar averaging expectation is that the BAO peak location 
through large voids should only be {\em slightly} greater than that 
for the best-fit FLRW model \citep{BuchCarf08,ROB13},
{i.e. difficult to detect.}
Moreover,
voids consist of deficits of galaxy densities rather than
excesses of galaxy densities, so
{procedures of}
finding and characterising voids 
{are more affected by Poisson noise
than in the case of} superclusters. Thus, the
relative stretching of the BAO peak location through voids
should be small compared to the best-fit FLRW model, and noisy.

\postrefereechanges{As a rough guide to what could
  be expected in the scalar averaging case, we 
  prepared an independent software package that implements the calculations illustrated in Fig.~2
  (left) of \citet{BuchRZA2} for the relativistic Zel'dovich
  approximation \citep{Kasai95,BuchRZA1} together with an early-epoch
  background model \citep{BuchRZA2}. We find that present-day
  effective (averaged) expansion factors $\aeff$ of typical (1$\sigma$
  in the first invariant of the extrinsic curvature tensor and zero in
  the second and third invariants; see \citealt{BKS00,BuchRZA2} for
  details) overdense or underdense regions at the 105{\hMpc} scale
  have effective scale factors of 0.95 or 1.04 times that of the
  extrapolated background model, i.e. the overdense region is about
  9\% compressed in comparison to the underdense region. The relative
  compression is the relevant quantity to consider here.}

\postrefereechanges{Instead of checking whether the BAO peak location through voids is stretched, a} more promising approach is to use superclusters. Lines of sight
passing through superclusters should pass through the 
compressed spatial regions, while lines of sight mostly passing far from the
superclusters should pass mostly through negatively curved space,
according to the scalar averaging models. 
\postrefereechanges{While superclusters are
unvirialised objects 
(typically of filamentary or spiderlike morphology;
e.g. \citealt{Einasto14filspider}), 
they} should not dominate the volume,
\postrefereechanges{so the metrical properties of the space
  they occupy are more likely to differ from
  volume-weighted mean quantities than in the case of voids.}
Thus, we 
expect to find a stronger and less noisy shift in the BAO peak location from superclusters than
from voids, and focus mostly on the former.

The BAO peak is well-determined in the Sloan Digital Sky Survey
Release 7 (SDSS~DR7) luminous red galaxy (LRG) sample
\citep{SDSSLRG01MNRAS}, with publicly available catalogues of both the
observational data and of ``random'' galaxies representing the
observational selection criteria. We use these to illustrate 
our method. We consider independent supercluster
catalogues from \citet{NadHot2013} and 
{\citet*{Liivamagi12}}, and
\citet{NadHot2013}'s Type 1 void catalogue.  The aim here is an
initial exploration of the method using real observational catalogues,
with the hope of detecting the BAO peak. The amplitude and sign 
of a shift in the BAO peak location or an upper limit to 
the shift should guide future work on observational catalogues
that are presently being prepared for public release or that will result from
the next generation of extragalactic telescope/instrument/survey
combinations. 

The observational catalogues are described in \SSS\ref{s-method-catalogues}.
The correlation function estimator is given in \SSS\ref{s-method-corrn}.
The calculation of an overlap 
between an LRG pair and a supercluster (or void) 
in flat FLRW comoving space 
is given in \SSS\ref{s-overlap-defn} 
[Fig.~\ref{f-overlap}, 
Eqs~(\ref{e-overlap-easiest}),
(\ref{e-overlap-general}), and
(\ref{e-overlap-defns})].
Determination of the BAO peak location
and its shift is described
in \SSS\ref{s-method-bao-locations}. 
Our method is conceptually simple but computationally 
heavy: numerical optimisation strategies are 
listed {in \SSS\ref{s-optimisation}.}

Results are presented in \SSS\ref{s-results} and discussed in
\SSS\ref{s-discuss}.  Conclusions are given in \SSS\ref{s-conclu}.
Distances and separations are stated in FLRW comoving units except if
stated otherwise, and $\Ommzero$ and $\OmLamzero$ are the
zero-redshift FLRW dimensionless matter density and dark energy
density parameters, respectively. In terms of common terminology, the
assumed comoving space is the FLRW space used 
{by, for example,}
\citet{NadHot2013}, i.e.  neither ``real space'' (since peculiar
velocities remain uncorrected for in the redshifts), nor ``redshift
space'' 
({in the sense that} 
$cz$ is not used as the radial distance).  The Hubble
constant is written $H_0 = 100 h$~km/s/{Mpc}.


\section{Method} \label{s-method}

\subsection{Observational catalogues}
\label{s-method-catalogues}

We use the ``bright'' and ``dim'' LRG samples 
{\em lrgbright}
{and} {\em lrgdim}
derived from the SDSS~DR7 \citep{SDSSLRG01MNRAS}, as provided by
\citet{Kazin2010} as observational and ``random'' catalogues for the
``Northern Cap
only''\footnote{\protect\url{http://cosmo.nyu.edu/~eak306/SDSS-LRG.html}}. 
These are converted to Euclidean comoving positions 
\begin{eqnarray}
  x\bggal_i &:=& 
  \rradialcomov\bggal_i \, \cos\alpha\bggal_i \,\cos\delta\bggal_i \nonumber \\
  y\bggal_i &:=& 
  \rradialcomov\bggal_i \, \sin\alpha\bggal_i \,\cos\delta\bggal_i \nonumber \\
  z\bggal_i &:=& 
  \rradialcomov\bggal_i \, \sin\delta\bggal_i ,
  \label{e-defn-xyz-bggal}
\end{eqnarray}
where $\rradialcomov\bggal_i$ is the radial comoving distance 
in {\hMpc} to the $i$-th LRG according to
an FLRW model with $(\Ommzero=0.32,\OmLamzero=0.68)$.

\tNobjects

\citet{NadHot2013}'s {\em v11.11.13} lists of superclusters identified
using the watershed algorithm in LRG samples {\em lrgbright}
{and} {\em lrgdim}
derived from the SDSS~DR7 \citep{SDSSLRG01MNRAS} are used
here\footnote{\protect\url{http://research.hip.fi/user/nadathur/download/dr7catalogue/}}.
These include 
the
right ascension and declination $\alpha\superclus_j,\delta\superclus_j,$ 
effective radius $\reffsuperclus_j$ in {\hMpc} and angular
radii 
$\theta\superclus_j$ of the superclusters\footnote{Columns 2, 3, 5, and 6, respectively, of the files
\filenamestyle{comovcoords/lrgdim/Type1clusters\_info.txt} and
\filenamestyle{comovcoords/lrgbright/Type1clusters\_info.txt}.},
which are converted to Euclidean comoving positions 
as in Eq.~(\ref{e-defn-xyz-bggal}), i.e.
\begin{eqnarray}
  x\superclus_j &:=& 
  \rradialcomov\superclus_j \, \cos\alpha\superclus_j \,\cos\delta\superclus_j \nonumber \\
  y\superclus_j &:=& 
  \rradialcomov\superclus_j \, \sin\alpha\superclus_j \,\cos\delta\superclus_j \nonumber \\
  z\superclus_j &:=& 
  \rradialcomov\superclus_j \, \sin\delta\superclus_j ,
  \label{e-defn-xyz-superclus}
\end{eqnarray}
where radial comoving distances 
{$\rradialcomov^{\mathrm{catalogue}}_j$} of the supercluster centres are 
derived as 
\begin{equation}
  {\rradialcomov^{\mathrm{catalogue}}_j}
  = \reffsuperclus_j/\tan\theta\superclus_j,
  \label{e-defn-rcomov-superclus}
\end{equation}
{the $\rradialcomov^{\mathrm{catalogue}}_j$ 
  values are}
converted to redshifts for the authors' FLRW model 
(with $\Ommzero=0.27,\OmLamzero=0.73, h=1.0$), and 
{the latter are} reconverted
to radial comoving distances 
{$\rradialcomov\superclus_j$}
for the FLRW model chosen in this work
$(\Ommzero=0.32,\OmLamzero=0.68)$.
\citet{NadHot2013}'s void catalogues for SDSS DR7 LRGs are extracted in the same 
way.\footnote{Files \filenamestyle{comovcoords/lrgdim/Type1voids\_info.txt} and
\filenamestyle{comovcoords/lrgbright/Type1voids\_info.txt}.}

\citet[][Fig.~4]{NadHot2013} used a conservative sky selection mask,
with a complicated sky geometry.
Superclusters outside of the main (``Northern Galactic Cap'') SDSS sky region
are excluded in this work
(i.e. superclusters with centres not satisfying 
$110^{\circ} < \alpha\superclus_j < 260^{\circ}$ are
ignored). 

\foverlap

\citet[][Fig.~1]{Liivamagi12} derived supercluster catalogues from the
SDSS~DR7, using a less stringent sky mask than \citet{NadHot2013}.
The celestial positions and FLRW radial distances of supercluster catalogue selected with an adaptive detection
threshold\footnote{Columns ``RA1deg'', ``DE1deg'', and ``Dist1'' of
\protect\href{http://cdsarc.u-strasbg.fr/viz-bin/nph-Cat/txt?J/A+A/539/A80/lrgadap.dat}{{\tt http://cdsarc.u-strasbg.fr/viz-bin/nph-Cat/txt?J/A+A/}}
\protect\href{http://cdsarc.u-strasbg.fr/viz-bin/nph-Cat/txt?J/A+A/539/A80/lrgadap.dat}{{\tt 539/A80/lrgadap.dat}},
representing estimates for the primary peaks of the superclusters.} 
are analysed here, since this catalogue is found by the
authors to approximate a volume-limited, distance-independent
catalogue. 
{As for the \citet{NadHot2013} catalogues, the}
catalogue FLRW radial distances are converted to redshifts
using $(\Ommzero=0.27,\OmLamzero=0.73, h=1.0)$ (\citealt{Liivamagi12}; {\SSS}~3.1)
in order to be consistent with those authors' analysis of the observations.
These inferred redshifts are then converted to radial comoving
distances { $\rradialcomov\superclus_j$}
for the FLRW model chosen for this work
$(\Ommzero=0.32,\OmLamzero=0.68)$.
{The Euclidean comoving positions follow from 
  Eq.~(\ref{e-defn-xyz-superclus}).}

Table~\ref{t-Nobjects} lists the numbers of real galaxies, 
simulated galaxies, real superclusters, and
real voids used in this work.

\subsection{Correlation function $\xi(s)$} 
\label{s-method-corrn}

The two-point auto-correlation function $\xi(s)$, where
$s$ is the pair separation in FLRW comoving space, is estimated
using the \citet{LandySz93} estimator 
{\citep*[see also][]{KerscherSzSz00}}
\begin{equation}
  \xi(s) = \frac{\DD(s)/N_{\DD} -2 \DR(s)/N_{\DR} + \RR(s)/N_{\RR}}{\RR(s)/N_{\RR}},
  \label{e-xi-defn-general}
\end{equation}
where $\DD(s)$, $\DR(s)$ and $\RR(s)$ are the 
data--data, data--random, and random--random
pair counts at $s$ for a given bin size $\Delta s = 10${\hMpc},
and $N_{\DD}$, $N_{\DR}$ and $N_{\RR}$ are the total
numbers of pairs. ``Data'' refers to observed 
{LRGs. ``Random''} refers
to a 
randomly resampled subset of a
simulated catalogue that mimics 
{the selection effects of the
real survey on the sky (boundaries and holes)
and in the radial direction}
{under the assumption that the intrinsic
  galaxy distribution is uniform in comoving space}
{\citep[][App.~A]{Kazin2010}}.

Here we use both data and random catalogues from \citet{Kazin2010}
(see \SSS\ref{s-method-catalogues}). As in
\citet[][App.~A]{Kazin2010}, we use radial 
{weights} designed to
optimise the signal at about 100{\hMpc} 
{\citep*{FeldmanKP94}}, as
provided in the catalogues, for both the data and random catalogues,
and the fibre collision {weight} is used for angular weighting in the
data catalogue.  Incompleteness is modelled by sampling the random
catalogue with a probability of success equal to the incompleteness
estimate for each simulated galaxy. 
The random catalogues have about 50 times as many galaxies as the
data catalogues. In practice {in this work},
the ratio of the requested number of random galaxies (prior to exclusion of some
objects due to incompleteness) to the number of data galaxies,
$\NRzero/N_{\mathrm{D}}$, has to be 
{several times} smaller than 50 in order
to achieve realistic computing times. This ratio is specified
per individual calculation below. 
Higher ratios reduce the contribution
of Poisson noise to the estimates of $\xi$.

\subsection{Overlap-dependent correlation functions}
\label{s-overlap-defn}

In order to determine the compression or stretching of comoving
features in the correlation function, we consider the subset of LRG
pairs that pass ``close'' to superclusters or voids, respectively. For
a given LRG pair, ``close'' is defined by considering the geometrical
overlap $\omega$ between the object and the pair, where the former 
is assumed to be the
interior of a 2-sphere and the latter a line segment, within flat
FLRW comoving space.  Figure~\ref{f-overlap} shows a galaxy--galaxy
pair close to a supercluster (or void). In this particular case, the overlap
is the chord length, i.e.
\begin{equation}
  \omega = 2\reffsuperclus \sin \theta.
  \label{e-overlap-easiest}
\end{equation}
The overlap $\omega$ is zero
when the impact factor is big,
i.e. when $\| \mathbf{e}\| > \reffsuperclus$,
where $\mathbf{e}$ is the normal from the 
LRG--LRG separation vector to the supercluster (void)
centre (see Fig.~\ref{f-overlap}).
 The most general case is given by 
\begin{equation}
  \begin{split}
    \omega = & \omega_{\mathrm a} + \omega_{\mathrm b} \; \mathrm{where} \\
    \omega_{\mathrm a} = & \left\{ 
    \begin{array}{ll}
      \min\left(\reffsuperclus \sin\theta, c_{\mathrm a}\right) &
      \mathrm{if} c_{\mathrm a} > 0 \\
      \max\left(-\reffsuperclus \sin\theta, c_{\mathrm a}\right) &
      \mathrm{if} c_{\mathrm a} < 0 \\
      0 & \mathrm{if} c_{\mathrm a} = 0
    \end{array} \right. \\
    \omega_{\mathrm b} = & \left\{ 
    \begin{array}{ll}
      \min\left(\reffsuperclus \sin\theta, c_{\mathrm b}\right) &
      \mathrm{if} c_{\mathrm b} > 0 \\
      \max\left(-\reffsuperclus \sin\theta, c_{\mathrm b}\right) &
      \mathrm{if} c_{\mathrm b} < 0 \\
      0 & \mathrm{if} c_{\mathrm b} = 0 {,}
    \end{array} \right.  \\
  \end{split}
  \label{e-overlap-general}
\end{equation}
where
\begin{equation}
  \begin{split}
    \hat{\mathbf{c}} := &
    \frac{\mathbf{c}}{\| \mathbf{c} \|}, \;\;
    c_{\mathrm a} :=  
    \overrightarrow{\mathrm{AS}} \cdot \hat{\mathbf{c}},\;\;
    c_{\mathrm b} :=  
    -\overrightarrow{\mathrm{BS}} \cdot \hat{\mathbf{c}}\\
    \mathbf{d} := & c_{\mathrm{a}} \hat{\mathbf{c}},\;\;
    \mathbf{e} :=  
    \overrightarrow{\mathrm{AS}} + \mathbf{d}.
  \end{split}
  \label{e-overlap-defns}
\end{equation}
The simple case illustrated in Fig.~\ref{f-overlap}  
and given by Eq.~(\ref{e-overlap-easiest}) occurs when
$c_{\mathrm a} > \reffsuperclus \sin\theta > 0$ and
$c_{\mathrm b} > \reffsuperclus \sin\theta > 0$.
      
For any given LRG pair, the overlap with each of the superclusters (or
voids) is calculated until either an overlap with $\omega \ge
1$~{\hMpc} is found or until all overlaps $\omega$ have
been calculated. An overlap is considered to occur in the
former case {and not in the latter.}
{For the calculation of $\xi$, the 
maximum value of $\omega$ for a given LRG pair is
forgotten once a decision has been made regarding 
the existence of an overlap for that pair.}

{For collecting overlap statistics
(\SSS\ref{s-disc-obs-caveats}, Table~\ref{t-npair-fractions}),
the calculation of overlaps continues without stopping
at the $\omega < 1$~{\hMpc} limit. This leads to much
longer calculation times, and high memory usage for
the calculation of quantiles. This 
calculation mode is not used for the main
calculations (of $\xi$).}

The expected differences between comoving correlation function
features in rigid comoving space versus inhomogeneous comoving space
are likely to depend on whether the pairs are mostly radial or mostly
tangential with respect to the observer. First, let us ignore
peculiar velocity effects (``redshift-space
distortion'', RSD, e.g. \citealt{BallingerPH96}). 
A compression or stretching
for superclusters or voids, respectively, can reasonably be expected
in either the radial or tangential direction 
in proportion to the ratio between
the locally averaged scale factor and the 
{(large-scale)}
mean effective 
scale factor $\aeff$.
In the tangential direction, a competing effect should
be expected from curvature. For example, positive curvature
would tend to act as a non-perturbative gravitational lens,
so that the would-be compressed BAO peak position is expanded.
However, given the estimated parameters of the
template metric presented in \citet{ROB13}, the
non-perturbative lensing effect would 
{most likely} be weaker
than the compression or stretching. 

Peculiar velocity effects 
\citep{Kaiser87}
complicate analysis of radial separations.
The BAO peak is likely to be more difficult to detect in the
radial direction.

Thus, we first divide pairs into whether or not they satisfy
the criterion $\omega \ge 1$~{\hMpc}, and secondly subdivide them
according to whether the LRG pair vector is closer to the line-of-sight
or rather closer to the sky plane, separating the two cases
at 45$^\circ$ using FLRW comoving space geometry. The overlap analysis
is carried out independently for $\DD(s)$, $\DR(s)$, and $\RR(s)$ pairs.
The \citet{LandySz93} estimator, Eq.~(\ref{e-xi-defn-general}), is 
used to estimate $\xi$ separately for each of these components, 
giving 
$\xiscrad$ and $\xisctang$, for the radial and tangential
overlapping components
of the correlation function for superclusters,
and 
$\xinonscrad$ and $\xinonsctang$, for the radial and tangential
non-overlapping components{,} and similarly 
$\xivoidrad$, $\xivoidtang$, 
$\xinonvoidrad$, $\xinonvoidtang$, 
for voids,
respectively.


\subsection{BAO peak locations} \label{s-method-bao-locations}

Since the aim of this work is to introduce a new observational
method of distinguishing the homogeneous and inhomogeneous 
assumptions for the comoving metric, we use simple methods that
are conservative in the sense that they are more likely to 
underestimate the shift in the BAO peak location 
{rather} than overestimate
it, and that we prefer to risk overestimating the error rather
than underestimating it.

\fxifull

\fxiNHsc

\fxiNHvoid

\fxiLiiva

Thus we use bootstraps 
{(\citealt{Efron79}; see also 
  \citealt{BarrowBhSon84boot})}
which tend to imply overestimates
of per-bin correlation function 
variances rather than underestimates \citep[e.g.][]{Snethlage00}.
Both the random
set of LRGs and the list of superclusters (or voids) are randomly
resampled (allowing repeats) for each bootstrap simulation.  The
observed LRG set is not resampled, since the aim is to test the
dependence of $\xi$ on the choice of 
{{\em subset}} of the observed LRG pairs.
Resampling the supercluster (or void) catalogue enables statistical
modelling of the sensitivity of the results to the inclusion or
exclusion of individual superclusters or voids, reducing the chance
that our main result---the shift in the BAO peak location---could 
be mostly the effect of statistical outliers
{resulting from the supercluster- or void-finding algorithms.}
Resampling the
random catalogue ensures that a major component of the uncertainty in
estimating the correlation function itself is 
statistically represented in the set of bootstrapped 
correlation functions.
The correlation function is calculated
separately for each bootstrap resampling.  
For illustrative purposes only,
the standard deviation at
each separation $s$ 
for the $\Nboot$ bootstraps of a given choice of LRG sample
and supercluster or void catalogue is calculated
and shown as error bars in plots of $\xi$.

Although we expect $\DD(s)$ to shift by $\Delta s \sim$ a few Mpc in
the scalar averaging case, to shift $\RR(s)$ by the same amount would
be difficult, especially given that we are trying to determine
the shift which has so far only been predicted (in this work)
qualitatively. The third type of pair count, $\DR(s)$, 
would also be difficult to model. Thus, we use the random catalogues
without modification, and do not expect a
shift of $\xi$ as simple as
$\xi'(s) = \xi(s + \Delta s)$. 
{In other words, calculating 
$\DD(s'), \DR(s')$ and $\RR(s')$, where $s'$ is 
a Lagrangian separation, in order to infer
$\xi(s')$, would not be easy.}
Instead, there is likely to be 
a general (smooth) change in shape. Since $\RR$ should be
a more or less smooth function, the BAO peak should still be visible
as a peak above the smooth ``background'' of the dominant component
of $\xi$. 

\fcubicfitfull

\fbaopeakfull

To separate the BAO peak from the main part of $\xi$, 
we fit a cubic $\xi_3$
to the smooth part of $\xi$ surrounding
the peak, i.e. in the range 
$40 \le s \le 180${\hMpc}, excluding 
$70 \le s \le 130${\hMpc}.
A linear or quadratic fit would 
be very sensitive to the choice of $s$ range for fitting;
a higher order polynomial than a cubic could fit part of
the peak, effectively weakening it.
We subtract the cubic fit, obtaining $\xi-\xi_3$,
and find the nonlinear least squares
best-fit Gaussian for the six 10{\hMpc} bins in the range
$70 \le s \le 130${\hMpc}, 
using
the Levenberg--Marquardt algorithm
\citep{Levenberg44,Marquardt63} 
implemented in {\sc leasqr} of
the optional {\sc octave} packet
{\sc octave-optim}.
Our template Gaussian is centred at $p_1$, has width $p_2$,
height $p_3$, and a vertical offset $p_4$, i.e.
\begin{equation}
  g(x) =  p_4 + \sqrt{2\pi} p_3 p_2 G(x,p_1,p_2),
  \label{e-gaussian-model}
\end{equation}
where $G(x,p_1,p_2)$ is a normal probability density
function of mean $p_1$ and standard deviation $p_2$.
{The} initial parameter guess is
\begin{equation}
  \begin{split}
    p_1 =& 100{\hMpcmath} \\
    p_2 =& \;\;5{\hMpcmath} \\
    p_3 =& \max_{70\hMpcmath \le s \le 130\hMpcmath}\{\xi-\xi_3\} \;- p_4 \\
    p_4 =& \min_{70\hMpcmath \le s \le 130\hMpcmath}\{\xi-\xi_3\}.
  \end{split}
\end{equation}
A weighting of $(1,2,4,4,2,1)$\footnote{The convention for this
  implementation of the Levenberg--Marquardt algorithm is that these
  {should normally} correspond to inverse
  standard deviations per bin; {thus, our
    weighting gives low statistical weight to the tails of the BAO
    peak and high statistical weight to a fixed BAO peak location 
    independent of the choice of subsample}.}  for the six bins, respectively, is applied in
order to reduce the influence of the tails of the peak, at the risk of
biasing the method towards finding the BAO peak location at the same
position in all cases. Since this is a bias towards underestimating
the true shift, this should give conservative results. We allow up to
30 iterations to find a fit. The centre of the Gaussian, $p_1$, is
considered to be the estimate $r$ of the BAO peak location.

While a Gaussian fit to the possibly shifted BAO peak is
a reasonable approach for this initial work, there is no
need to assume that the errors in the fit themselves follow
a Gaussian distribution. Since we have $\Nboot$ 
bootstrap estimates of the function $\xi$ in any given case, 
we calculate the shift of the BAO
peak location between a pair of cases $r', r''$, 
by using the sets of BAO peak 
locations for the bootstraps as discretised estimates of their 
probability density functions. This avoids 
{assumptions about the shapes of the} 
probability distributions of the two {estimates.}
That is, 
for each estimate of 
$\ralltang - \rsctang,$
$\ralltang - \rvoidtang$,
$\rnonsctang - \rsctang,$ or
$\rnonvoidtang - \rvoidtang$, 
we take $r'$ and $r''$ from {bootstrap realisations
for the two cases, respectively.}
Non-convergent BAO peak location estimates are ignored and
noted in the table of results.
For the 
{estimates of $\rnonsctang - \rsctang$ and
$\rnonvoidtang - \rvoidtang$,} 
the two estimates in a pair
come from the same bootstrap realisation.
Due to long calculation times, $\Nboot$ for the 
\citet{Liivamagi12} case is less than {$\Nboot$} for the full sample
(without supercluster overlap detection), so half of the full
sample realisations are ignored for 
$\ralltang - \rsctang$ in this case.
We use robust statistics of this set of realisations 
to estimate the shift,
i.e. we calculate the median
$\mu(r'-r'')$
and we use
$1.4826$ times the 
median absolute deviation \citep{Hampel74}
as an estimate 
of the standard deviation
$\sigma(r'-r'')$.
(Thus, we do not use variances per bin from the bootstraps,
nor do we assume Gaussian error distributions in the 
estimates of the peak locations.)

\fbaopeakNHsc

\fcdfNHsc

\subsection{Optimisation of calculation speed}
\label{s-optimisation}

The calculation time of the two-point auto-correlation functions 
is dominated by $N_{\mathrm R}^2$, but is also roughly proportional
to the number $N\superclus$ of superclusters or voids. 
Thus, it scales as $N_{\mathrm R}^2 N\superclus$, requiring 
many cpu hours of computation.
{Optimisations used in the present work}
include:
\begin{list}{(\roman{enumi})}{\usecounter{enumi}}
  \item
    while counting pairs in parallel ({\sc openmp}) threads, 
    store the binned pair counts per 
    outer-loop galaxy (the outer loop is parallelised)
    and only sum these (per separation bin) after the 
    threads have finished;
    this requires a modest amount of extra memory but favours speed
    by avoiding atomic/critical instructions;
  \item
    for a given galaxy--galaxy pair, stop calculating overlaps with
    superclusters or voids if an overlap greater than the threshold
    (1{\hMpc} in this work) has already been found (as stated above);
  \item 
    inline the vector-related functions and the function to calculate
    overlap, Eq.~(\ref{e-overlap-general}), by their inclusion in the
    same file as that for pair count functions and using the {\sc gcc}
    compile-time optimisation option {\tt -O3}.
\end{list}
{Significant speedup for flat space calculations
  \postrefereechanges{might} also be possible using a $k$d tree approach
  \citep[e.g.][]{Moore01kdtree}.}

\section{Results} \label{s-results}

Figure~\ref{f-xi-full} shows a sharp BAO peak for the bright sample
and a broad, less well-defined peak for the dim sample. Thus,
we analyse the former.
The calculations described below, i.e. for the bright sample,
made using the optimisations described in \SSS\ref{s-optimisation}, 
represent about 240,000 cpu-core--hours of
computation on Intel Xeon E7-8837 processors.

The correlation functions for tangential pairs either overlapping or not
overlapping with \citet{NadHot2013} superclusters or voids
are shown as the two upper curves (red, purple online) 
in Figs~\ref{f-xi-NH-sc} and ~\ref{f-xi-NH-void}. Keeping in mind
that the per-bin standard deviation (estimated from bootstraps as shown or by
other methods) does not directly show the uncertainty in the existence or
position of the BAO peak, the best estimate of $\xi$ does appear to be
present for all four {of these curves} in 
the separation bins centred at either 95{\hMpc} or 
105{\hMpc}. For overlaps with 
superclusters or for LRG pairs not overlapping with voids 
the peak appears to be located at 95{\hMpc}. For overlaps with voids
or for non-overlaps with superclusters, the BAO peak appears to
be at 105{\hMpc}, i.e. at the same location as that of the full sample.

\fbaopeakNHvoid

\fcdfNHvoid

\fbaopeakLiiva

\fcdfLiiva

The correlation functions for radial pairs either overlapping or not
overlapping with \citet{NadHot2013} superclusters or voids
are shown as the two lower curves (blue, green online) 
in Figs~\ref{f-xi-NH-sc} and ~\ref{f-xi-NH-void}. Clearly,
the Kaiser effect 
\citep[large-scale smooth infall][]{Kaiser87} 
significantly affects $\xi$ on these scales.
Moreover, since we define the split between radial and tangential pairs at an 
angle of 45$^\circ$ with respect to the plane of the sky, there
are many fewer radial than tangential pairs (the sky plane is two-dimensional).
{Thus, use of the radial case for detecting a shift 
in the BAO peak location would risk ambiguity in identifying the peak
and most likely be subject to higher systematic error than the tangential
case. The radial curves} are not used further in the present work.

The \citet{Liivamagi12} supercluster correlation functions are shown
in Fig.~\ref{f-xi-Liiva}.  There is a great difference between the
tangential and radial supercluster-overlapping functions, and a very
strong difference between these and {the}
non--supercluster-overlapping pair correlation functions. This is
easily interpreted as an effect of the integral constraint
{(see, e.g., Fig.~7, lower panel, 
  \citealt{RoukP94})},
for which
we have not made any corrections, since our aim is to detect the shift
in the BAO peak location rather than attempting to model the general
change in the shape of the correlation function.  A BAO peak for the
\citet{Liivamagi12} tangential supercluster-overlapping case (red
thick curve) is barely visible by inspection of Fig.~\ref{f-xi-Liiva}.
In the non--supercluster-overlapping case, a BAO peak is obvious in
the 100--110{\hMpc} bin, but a broad weaker peak with maxima in the
70--80{\hMpc} and 80--90{\hMpc} bins suggests that the detection is
ambiguous.  However, we retain the method defined above
(\SSS\ref{s-method-bao-locations}), without any modification, in order
to examine these two curves, i.e. for each of the bootstrap estimates
of these correlation functions, we subtract a best-fit cubic and
least-squares fit a gaussian in order to estimate the BAO peak
location. As shown below, subtracting best-fit cubics yields clear
BAO peaks in both cases.

\tkolsmirnov

\tcompression

\tnpairfractions

{Cubic fits for individual
bootstrap correlation functions, 
as described in \SSS\ref{s-method-bao-locations},
are shown in Fig.~\ref{f-cubicfit-full}.
Clearly, the amplitude of the BAO peak after subtracting
the cubic would differ significantly from that obtained
from fitting a smooth function with a narrower exclusion region
around the peak. Since the aim here is to measure the shift
in the peak location, not the peak's amplitude, this should not
affect our results.
Cubic-subtracted bootstrap correlation functions}
are shown in Figs.~\ref{f-baopeak-full},
\ref{f-baopeak-NH-sc},
\ref{f-baopeak-NH-void}, and
\ref{f-baopeak-Liiva},
for the full sample and 
for the tangential LRG
pairs.
The BAO peak is clearly visible in all of these plots,
although some of the correlation functions for individual
bootstraps fail to show it clearly.

The upper panels of Figs~\ref{f-baopeak-NH-sc} and
\ref{f-baopeak-Liiva}, i.e. for LRG--LRG pairs overlapping
superclusters, clearly show that most \citep{NadHot2013} or many
\citep{Liivamagi12} of the bootstrapped correlation function BAO peaks
are centred in the 90--100{\hMpc} bin. In contrast, the lower panels
of these two figures show similar behaviour to
Fig.~\ref{f-baopeak-full}: the BAO peak is centred in the
100--110{\hMpc} bin. The lower panel of Fig.~\ref{f-baopeak-NH-void},
for LRG--LRG pairs not overlapping with voids, shows a weaker
compression; a minority of the bootstrap correlation functions 
have a BAO peak centre in the 90--100{\hMpc} bin.

Fitting Gaussians to these bootstrapped BAO peaks 
yields estimates of the 
BAO peak locations. 
The cumulative distribution functions (cdf's) of these
estimates are shown in Figs.~\ref{f-cdf-NH-sc},
\ref{f-cdf-NH-void}, and
\ref{f-cdf-Liiva}. It is obvious that the pairs of cdf's are
incompatible in both 
Figs.~\ref{f-cdf-NH-sc} and
\ref{f-cdf-Liiva}, i.e. when the supercluster-overlapping BAO peak
location is compared to either the full sample BAO peak
location or the complementary-pair BAO peak location.
The Kolmogorov--Smirnov two-sided probabilities that
the two members of a pair of cdf's represent samples drawn from
a single continuous underlying distribution are listed
in Table~\ref{t-kol-smirnov}. For the supercluster-overlapping 
comparisons (first and second columns of Table~\ref{t-kol-smirnov}), 
identity of the distributions is rejected
to very high significance (greater than 99.99\%).

The probabilities in Table~\ref{t-kol-smirnov} are not
sufficient to show that the differences in distributions
constitute shifts in the 
BAO peak location rather than, for example, differences
in noise levels, i.e. widths rather than central tendencies.
The best estimates of the shifts are listed 
in Table~\ref{t-compression}.


Table~\ref{t-compression} shows that the supercluster-overlapping
BAO peak location is about 6--7{\hMpc} less than that of the complementary set
of LRG pairs and
about 4{\hMpc} less than that of 
the full sample of pairs. 
{These shifts are} statistically 
significant at about the $2.5\sigma$ level 
(in terms of Gaussian intuition)
for the \citet{NadHot2013}
superclusters, and at weaker significance for the \citet{Liivamagi12}
superclusters.

\section{Discussion} \label{s-discuss}

Figures~\ref{f-baopeak-full}--\ref{f-cdf-NH-sc},
\ref{f-baopeak-Liiva}, and \ref{f-cdf-Liiva}, and 
Tables~\ref{t-kol-smirnov} and \ref{t-compression} clearly
show that the supercluster-overlapping
BAO peak location for tangential pairs 
is about {6--7}{\hMpc} less than that for the non--supercluster-overlapping pairs, i.e. 
superclusters correspond to 
a compression {of about 6\%. 
A weaker} compression is found in comparison to the 
{full-sample} tangential pairs,
{i.e.} without overlap selection.
The best estimates of the shift in 
the void-overlapping {case} are much
{smaller, 
as expected for a void-dominated best-fit metric,
though with high uncertainties.
This trend of strong relative compression of
the supercluster-overlapping BAO peak location
and a weak stretching in the void-overlapping case
qualitatively agrees with 
{what is expected for inhomogeneous models.}
Analysis of other surveys should reduce the
statistical uncertainties to see if the
trend continues to favour 
{an inhomogeneous metric}.}

{Theoretical interpretation of these 
results will require taking} into account the typical 
fraction of a separation path that actually {overlaps} with
a supercluster or void. For separations in the $70 < s/{\hMpc} < 130$
range, the median overlap is about $70, 85,$ and $40${\hMpc} for
the \citet{NadHot2013} superclusters, voids, and 
\citet{Liivamagi12} superclusters, respectively,
{i.e. roughly 70\%, 85\%, and 40\% of the
BAO-scale separation paths should be compressed, stretched, and
compressed, respectively.
We present
observational caveats within the FLRW (rigid comoving) interpretation 
of the data} in \SSS\ref{s-disc-obs-caveats}.
{In \SSS\ref{s-disc-scalar-av-improve}, we
  propose a method} of more accurate analysis by using an
effective template metric rather than the $\Lambda$CDM metric for the
{assumed cosmological model.
In \SSS\ref{s-disc-other-inhomog},}
we discuss possible relations of 
{the present work} to
other well-known dark-energy--free general-relativistic
approaches to cosmology.
{We focus on}
observational methods or claims of
detecting metric inhomogeneity
{in \SSS\ref{s-disc-inhomog-claims}.}
{In \SSS\ref{s-disc-other-LCDM-rejections},
we list some observational results that reject the $\Lambda$CDM model.}

\subsection{Observational caveats and improvements} 
\label{s-disc-obs-caveats}

{Could the} superclusters listed by \citet{NadHot2013} and
\citet{Liivamagi12} constitute very rare overdensity fluctuations, so 
that the LRG pairs that overlap them
(given our definition in \SSS\ref{s-overlap-defn})
constitute a strongly biased subset that favours rare, highly 
nonlinear 100{\hMpc} fluctuations? 
Table~\ref{t-npair-fractions} shows that on the contrary, 
the overlapping pair fractions are high, about 80--90\%.
The numbers of superclusters vary by a factor of 10 between
the two groups' analyses
(Table~\ref{t-Nobjects}), but in neither case can 
the compressed BAO scale be attributed to the rarity of the
superclusters and associated LRG--LRG pairs.

Our result could be thought of in terms of a ``non-random''
jackknife type analysis of the set of $\sim 10^9$ SDSS DR7 
``bright'' LRG--LRG pairs. Given that the full set of $\sim 10^9$
pairs gives the standard BAO peak location, how easy is it
to choose about 10--20\% of these pairs to ignore so that the
BAO peak location of the remainder is reduced by 
about 6{\hMpc}? We know one answer (for the tangential pairs): 
choose either the
\citet{NadHot2013} or \citet{Liivamagi12} supercluster
catalogue, and ignore those pairs that
do not overlap these superclusters by 1{\hMpc} or more. 
Only a minority of pairs are excluded, and the BAO peak 
standard ruler is compressed by about 6{\hMpc}.
Our result would be difficult to explain 
{as a} bias towards very rare fluctuations.

Could strong or weak gravitational lensing create a bias
that has been ignored here?
Gravitational lensing, in the sense of observations interpreted
by perturbative calculations {against} an FLRW background, does not imply
large offsets in path lengths projected to a comoving spatial slice.
For example, multiply imaged galaxies 
have time delays ranging from a few days to a few years
\citep[e.g. Table 2][]{Paraficz10},
i.e. radial paths should vary by sub-parsec scales rather than
megaparsec scales. In the tangential direction, strong lensing
typically plays a role on arcsecond scales, and weak lensing
(shear of galaxy images) shows correlations at arcminute 
scales \citep[e.g.][and refs therein]{Schneider05}. These
are both much smaller than the degree scales that would
need to affect the scale of the BAO peak in tangential directions.


Could peculiar velocity effects be relevant? We chose to use
the ``tangential'' pairs, i.e. those which lie within 45$^\circ$
of the sky plane, so peculiar velocity effects should be weak. 
Within a $\Lambda$CDM model, a (radial) peculiar velocity of 1000 km/s
at $z = 0.3$ consists of a multiplicative error in $(1+z)$ of $1.003$,
corresponding to about 10{\hMpc}. 
The ``finger of God'' effect is unlikely to be significant,
since we consider separations well above 10{\hMpc}. 
However, the larger scale Kaiser effect \citep{Kaiser87} is of the
order of magnitude to potentially be of concern.
This has long been expected to 
have significant effects in the radial direction on 
redshift-space separation scales of tens of megaparsecs,
in the sense of shifting power to smaller scales
\citep[e.g.][]{BallingerPH96,MatsubaraSuto96}.
Recent models and estimates 
{\citep[e.g.][and references therein]{SongSOOL14}}
show considerable differences between the overall shapes of 
the radial and tangential
correlation functions, qualitatively consistent with what
is shown in the figures above.
\citet{Jeong14BAOradvstan} estimate that the BAO peak location
estimated directly from $\xi$, without removal of the smooth component
of the function, can yield a difference between the {exactly} radial and
tangential directions of about 3{\hMpc}.
However, they also estimate
\citep[Fig.~6 right,][]{Jeong14BAOradvstan} that when the smooth
component is removed (by two methods different to our choice of 
subtracting a best-fit cubic), the difference is reduced to 
about $0.2$--$1.0${\hMpc}.

In our analysis, 
{we do not compare exactly radial pairs to exactly
tangential pairs; we compare ``tangential'' samples to each other,
defined by a 45$^\circ$ split. Nevertheless, let us}
consider an extreme case in order
to estimate an upper bound to this effect.
If one of our ``tangential'' samples would
consist of exactly line-of-sight pairs and the other uniquely of 
45$^\circ$ pairs, then the \citet{Jeong14BAOradvstan} analysis
for smooth-component--removed BAO peak locations would suggest
a difference
{below their $0.2$--$1.0${\hMpc} estimate
  for exactly tangential versus radial pairs.}
{Our BAO peak location estimation method,
which is preceded by the removal of a best-fit cubic,
would be} likely to give a similar result.

More realistically, 
{any two subsamples that we compare in this
work consist
of pairs whose angles with respect to the sky plane are
distributed between 0$^\circ$ and 45$^\circ$, not
Dirac-delta distributed at one or the other of these.
However, the distributions are unlikely to be exactly identical.
It} is likely that, at least {due to the role
  of} Poisson noise, either the median angle to the line-of-sight of
the ``tangential'' pairs that 
overlap superclusters is slightly greater than the median angle of
pairs that do not overlap superclusters, or vice-versa. 
{This} angular difference, or 
{the overall distributions of angles with respect to the sky plane,}
should give a {small} distortion in our result, {but} by 
{much less than the} $ 0.2$--$1.0${\hMpc}
{indicated above.}
{Correcting for this}
would not remove the {shift that we have detected}
for the \citet{NadHot2013} and \citet{Liivamagi12} 
{catalogues, as} given in
Table~\ref{t-compression}, but would be useful to include in future work 
in order to obtain more results at the sub-megaparsec level.

The present results should be obtainable at higher statistical
significance in existing catalogues (e.g., BOSS DR11 LOWZ and CMASS) and future data
sets such as those from 
{SDSS-III/BOSS DR12, SDSS-IV/eBOSS,}
{LSST, EUCLID, 
VISTA/4MOST \citep{deJong12VISTA4MOST} and
DESI \citep{Levi13DESI}.}
At lower redshifts than
those of SDSS DR7 LRGs ($z\sim 0.3$), the Universe is more inhomogeneous,
but there is very little volume to sample well at the 100{\hMpc} scale.
At higher redshifts $z < 1$, the amount of volume to sample is higher,
but the inhomogeneity (virialisation fraction) is lower, so the effect
should be weaker. 
{Despite these difficulties,
the} redshift evolution of the BAO peak location {shift}
for superclusters,
{i.e. the  $\rnonsctang - \rsctang$ vs $z$ relation,}
should constitute a new statistic that would
need to be explained quantitatively in scalar averaging, 
dark-energy--free models or in $\Lambda$CDM or other 
FLRW models, e.g. with inhomogeneous dark energy.
{As observational accuracy
improves and theoretical predictions are made, 
this will provide {a geometrical} test to
distinguish the two.}

\subsection{Improved analysis assuming scalar averaging}
\label{s-disc-scalar-av-improve}

The $\Lambda$CDM best-fit metric interpretation of the data,
which we have used above,
uses a distance--redshift relation based on the assumption
that the expansion rate of the averaged 3-spatial slices
exactly equals that of homogeneous, unaveraged 3-spatial slices.
This assumption, of the FLRW models in general,
is that the implicitly 
averaged constant-curvature time slices happen, fortuitously,
to evolve in the way that idealised exactly homogeneous slices
would evolve.
A more realistic interpretation would be possible using
a template metric \citep{Larena09template,ROB13}.
This would not only help test the validity of the template metrics
that have so far been proposed, but should also 
yield more accurate estimates of the BAO peak location shift.

Theoretical work in inhomogeneous cosmology in general
should also be useful not only in making
predictions and interpretations, but also in improving 
the observational analysis.
For example, we have assumed that the projection of photon
paths from the past time cone to the comoving spatial section
does not significantly affect the expected compression or
stretching of comoving separations. Since we focus on 
tangential separations, this seems to be a reasonable
assumption, but ray tracing in scalar averaging
models or Swiss cheese models 
(e.g. \citealt{FutamaseSasaki89,
  Marra07};
{\citealt*{Szybka11raytrace,
    Fleury13multigeom,
    Fleury13Hvariance}})
should be checked, especially for extending the method
used here to radial pairs. 
Changes to the BAO peak location in the radial direction
are likely to be more complicated to extract than for tangential
pairs, but might contain more useful cosmological information.

\subsection{{Other inhomogeneous approaches}}
\label{s-disc-other-inhomog}

{Other dark-energy--free general-relativistic
approaches to observational cosmology 
include Stephani models  \citep{DabHend98} and
Lema{\^\i}tre--Tolman--Bondi (LTB) solutions.
The latter are often applied as a local void model
\citep[e.g.][]{MustaphaHE97,
Celer99,
Tomita01void,
Alnes05voidmodel,
GBH08Gpc2p5,
Enqvist08LTBHpec,
AlexBiswas09void250Mpc,
Biswas10,
HuntSarkar10,
BCK11review},
sometimes infer a local 
hump from observational data 
{\citep*{CBK10,KolbLamb09},} 
and are widely used \postrefereechanges{in}
``Swiss cheese'' models
\citep[e.g.][and references therein]{BiswasN08Swiss,BolejkoC10SwissSzek,Lavinto13}.
\postrefereechanges{LTB models are also used to argue that
  within homogeneous (FLRW) 
  cosmology for non-evolving dark energy, i.e.
  a cosmological constant model, the value of the cosmological
  constant will be mis-estimated by about a percent if
  non-perturbative calculations are not taken into
  account \citep{Romano11,RomanoNSS14}.}
Independently of the kinematic Sunyaev--Zel'dovich effect
arguments against the local void model
{\citep*{MossZS11defeat,ZhStebbins11void1Gpc,ZibinMoss11voidkSZ},}
a local void (or hump) should not have a strong effect 
on the BAO peak location test introduced here, since our
test is differential,
{comparing different subsamples of a
  single survey in a given volume of space.}
However, if the whole survey (e.g. SDSS DR7
LRGs) were contained in a 1{\hGpc} void centred not too
far from the Galaxy, then the large-scale gradient in underdensity
could probably introduce a small bias, in a similar way in which
it would affect estimates of the ``full'' BAO peak location.

Swiss cheese models typically use an FLRW ``background'' model
for the ``cheese'' and LTB models for the ``holes''.
{For a given Swiss cheese model, it}
should be possible to study the differential evolution of
pairs of tracer galaxies that overlap cheese versus the
complementary set of pairs, 
or that overlap voids versus the complement. Since most tracers
should lie in the ``cheese'', their pairs would already be
comoving by definition of the model. 
{It should be possible to 
calculate the pairs' comoving separations by integration of 
the (analytically exact) {metric} of the model.}
A differential study of the complementary sets of pairs
would probably provide the simplest way to model the
expected shift in the BAO peak location in this class of 
inhomogeneous 
{models}.}

\subsection{Observational methods 
  {of detecting metric inhomogeneity
    and claims of detections}}
\label{s-disc-inhomog-claims}

{Observational methods of distinguishing
the literally homogeneous FLRW model from statistically
homogeneous relativistic models have mostly focussed
on the distance-modulus--redshift relation (to fit
supernovae type Ia observations) or 
estimates of $H(z)$ versus $z$, i.e. the expansion-rate--redshift relation
\citep[e.g.][]{SmaleWilt11SNe,BoehmRasan13}.
Perturbed FLRW strategies for detecting inhomogeneity
also focus on the expansion rate 
\citep[e.g.][]{Rasanen12pertangdiam}, and especially its variance
\citep{BDDurrerMSchwarz14}.
[Redshift drift is a test that distinguishes spherically symmetric inhomogeneous
models, the LTB and Stephani models, from each other and from 
the $\Lambda$CDM model \citep[e.g.][]{BalcDab13reddrift}.]}

{While several 
dark-energy--free inhomogeneous models
claim to have fit several sets of extragalactic observations
(\SSS\ref{s-disc-other-inhomog}; see also 
\citealt{DuleyWilt13}), some focus on tests that
are qualitatively new.
\citet{Fleury13multigeom} argue that a single FLRW metric is 
relativistically inaccurate for modelling both wide-angle
observations [cosmic microwave background (CMB), BAO] 
and narrow-angle observations
(type Ia supernovae), and appear to resolve conflicting
estimates of \postrefereechanges{$\Ommzero$} using a Swiss cheese model with FLRW cheese.}
On scales up to tens of megaparsecs from the Galaxy,
\citet{Wiltshire12Hflow} find that galaxy peculiar velocity
flow analyses imply a relation between the
Local Group rest frame and 
{what is normally considered to 
be the CMB comoving 
rest frame that is different from
the FLRW expectation}.
\citet{Saulder12anisoH} propose to analyse the dependence
of the expansion rate (Hubble parameter) on 
line-of-sight environment, i.e. mostly through dense structures
versus mostly through voids and present preliminary results.

\subsection{{Observational methods that reject the $\Lambda$CDM model}}
\label{s-disc-other-LCDM-rejections}

{The method presented here is not the only one that potentially or
  already rejects the $\Lambda$CDM model.  While many observations
  agree with the latter, several recent observational results (other
  than the present work) reject it.  An incomplete list includes the
  following.}  \citet{Flender13ISW3sig} find that {what is generally
  accepted as a detection of the} integrated Sachs--Wolfe (ISW) effect
in the CMB is stronger than the $\Lambda$CDM expectation at the
3$\sigma$ level.  {\citet*{WiegBO14}} find that Minkowski functional
analysis {(which implicitly includes all orders of $n$-point
  auto-correlation functions)} of the SDSS DR7 significantly rejects
the $\Lambda$CDM model on scales of several tens of {megaparsecs in
  volume-limited samples of 500{\hMpc} (3$\sigma$) and 700{\hMpc}
  (2$\sigma$).}  
\postrefereechanges{Earlier Minkowski functional analysis
on smaller and sparser catalogues, using the same
or complementary methods, had found that the 
fluctuations were compatible with $\Lambda$CDM mock
catalogues \citep[e.g.][]{Kerscher01Mink,Hikage03Mink}.}
\citet[Fig.~6,][]{ChuangBOSSLCDMruledout14} find that
the CMASS and WiggleZ \citep{Blake12WiggleZjoint} normalised growth
rate estimates contradict the Planck Surveyor cosmic microwave
background $\Lambda$CDM model at about 2$\sigma$ significance.  By
comparing CFHTLenS weak lensing analysis with a parametrisation of the
$\Lambda$CDM version of the FLRW model using Planck and WMAP data,
\citet{MacCrannJain14} reject the model at the 90--96\% level (but at
only the 64\% level if a sterile neutrino is {included}).
{\citet*{Battye14fivesigmaantiLCDM}} reject the $\Lambda$CDM model at
the 5$\sigma$ level {based} on contradictions between large-scale and
small-scale power, unless the sum of neutrino masses is increased
above that of the Planck base model of 0.06~eV
\citep[Sect.~6.3.1,][]{PlanckXVIcosmoparam13}.
\postrefereechanges{The BAO peak (without any shift measurement)
  detected in the Lyman $\alpha$ forest in front of about $10^5$
  quasars at $z \sim 3$ was estimated to reject the $\Lambda$CDM model
  at the 2.5$\sigma$ level \citep{Delubac14BAOLyaforest}.  A
  significant contradiction exists between Galactic metal-poor star
  estimates of the pre-Galactic ${}^7$Li abundance and the
  abundance inferred from either primordial big bang nucleosynthesis
  or the cosmic microwave background interpreted according to an FLRW model
  \citep{Cyburt08Li7}, unless new particles such as decaying
  gravitinos are assumed \citep[e.g.][]{Cyburt13Li7gravitino}.}

\section{Conclusion} \label{s-conclu}

We have introduced supercluster-overlap dependent BAO peak location
estimation as a new observational method of distinguishing the FLRW
models (including the $\Lambda$CDM model), which assume rigidity of
comoving space, from scalar averaging models, which allow comoving
space to be curved and compressed or stretched 
by structure formation. The initial results from
the SDSS DR7 are promising, showing that by choosing the sharpest
signal, that of ``tangential'' LRG pairs, a detection of compression
in the BAO peak location for supercluster-overlapping pairs versus
complementary pairs is significant at about the 2.5$\sigma$ level for
two different supercluster catalogues: $6.6\pm2.8${\hMpc} for 
{\citet{NadHot2013}} 
superclusters and $6.3\pm2.6${\hMpc} for
\citet{Liivamagi12} superclusters.  Compression relative to the full
(tangential) sample is, as expected, {weaker:} $4.3\pm1.6${\hMpc} and
$3.7\pm2.9${\hMpc}, respectively.  Stretching in the void-overlapping
case is numerically consistent with what is expected (negative
compression in columns 3 and 4 of Table~\ref{t-compression},
{i.e. stretching}), but
statistically insignificant.  The differences in the bootstrap
estimates of the BAO peak locations for supercluster-overlapping pairs
versus complementary pairs are strikingly obvious in
Fig.~\ref{f-baopeak-NH-sc} and Fig.~\ref{f-cdf-NH-sc} (lower panel)
for the \citet{NadHot2013} superclusters, and in
Fig.~\ref{f-baopeak-Liiva} and Fig.~\ref{f-cdf-Liiva} (lower panel)
for the \citet{Liivamagi12} superclusters. The corresponding
Kolmogorov--Smirnov formal estimates of incompatibility in the
cumulative distribution functions (Table~\ref{t-kol-smirnov}) reflect
the strength of the differences visible by inspection in these
figures.

{Qualitatively, these results are 
consistent with what is 
expected
from scalar averaging and inconsistent with what is expected
in the rigid comoving space models, including 
{the}
$\Lambda$CDM model. 
Theoretical work on the BAO peak location shift
in both approaches, together with
observational development of the test, 
may potentially challenge the $\Lambda$CDM model
and constrain backreaction models.}
\postrefereechanges{For example, application of our test to 
$\Lambda$CDM $N$-body simulations would provide a check
in terms of a widely used tool of FLRW cosmology.
This test should be compared to the expected 
$\Lambda$CDM
low-redshift BAO peak location shift
of less than 0.3\% 
[Eqs~(30), (31), \citealt*{SherwZald12peakscale}; see also
\citealt*{McCullagh13peaklocation}; and 
\postrefereechangesbis{references therein; 
  see also e.g. \citet{SlepianEis15}, for 
   baryon--dark-matter relative-velocity corrections}]
which is much smaller than the 
\postrefereechangesbis{environment-dependent} 
shift of 6\% which is found
here.}

\section*{Acknowledgments}

Thank you to Tomasz Kazimierczak, Hirokazu Fujii,
St\'ephane Colombi,
{Alexander Wiegand, Bartosz Lew,}
{Martin Kerscher,}
Istv\'an Szapudi, Pierre Astier, 
{Pierre Fleury,
\postrefereechangesbis{Zachary Slepian}
 and an anonymous referee}
for useful 
comments related to this work.
Part of this work consists of research conducted 
{within} the scope of the HECOLS International Associated Laboratory,
{supported in part by the Polish NCN grant DEC-2013/08/M/ST9/00664.}
\postrefereechangesbis{Part of this 
  project was performed under grant 2014/13/B/ST9/00845 of
  the National Science Centre, Poland.}
BFR thanks
the Centre de Recherche Astrophysique de Lyon for a warm welcome
and scientifically productive hospitality.
A part of this project has made use of 
{computations made under} grant 197 of the
Pozna\'n Supercomputing and Networking Center (PSNC).
{Some of JJO's contributions to this work
were supported by the Polish Ministry of Science and Higher Education
under ``Mobilno\'s\'c Plus II edycja''.}
A part of this work was conducted within the ``Lyon Institute of
Origins'' under grant ANR-10-LABX-66.
Use was made of the Centre de Donn\'ees astronomiques de Strasbourg 
(\url{http://cdsads.u-strasbg.fr}),
the GNU {\sc plotutils} graphics package,  
and
the GNU {\sc Octave} command-line, high-level numerical computation software 
(\url{http://www.gnu.org/software/octave}). 
%

Funding for the SDSS and SDSS-II has been provided by the Alfred
P. Sloan Foundation, the Participating Institutions, the National
Science Foundation, the U.S. Department of Energy, the National
Aeronautics and Space Administration, the Japanese Monbukagakusho, the
Max Planck Society, and the Higher Education Funding Council for
England. The SDSS Web Site is \url{http://www.sdss.org/}.
The SDSS is managed by the Astrophysical Research Consortium for the
Participating Institutions. The Participating Institutions are the
American Museum of Natural History, Astrophysical Institute Potsdam,
University of Basel, University of Cambridge, Case Western Reserve
University, University of Chicago, Drexel University, Fermilab, the
Institute for Advanced Study, the Japan Participation Group, Johns
Hopkins University, the Joint Institute for Nuclear Astrophysics, the
Kavli Institute for Particle Astrophysics and Cosmology, the Korean
Scientist Group, the Chinese Academy of Sciences (LAMOST), Los Alamos
National Laboratory, the Max-Planck-Institute for Astronomy (MPIA),
the Max-Planck-Institute for Astrophysics (MPA), New Mexico State
University, Ohio State University, University of Pittsburgh,
University of Portsmouth, Princeton University, the United States
Naval Observatory, and the University of Washington.


%
%


\subm{ \clearpage }

%




\end{document}